\documentclass[aps,prb,10pt,twocolumn,a4paper,  notitlepage, longbibliography]{revtex4-1}

\usepackage{amsmath}
\usepackage{amssymb}
\usepackage{upgreek}
\usepackage{units}
\usepackage{booktabs}
\usepackage{graphicx}

\usepackage{hyperref}
\hypersetup{colorlinks=true, citecolor=blue, urlcolor=blue, linkcolor=blue}

\usepackage{xcolor}

\begin{document}
\title{Chiral Hinge Magnons in Second-Order Topological Magnon Insulators}

\author{Alexander Mook}
\affiliation{Department of Physics, University of Basel, Klingelbergstrasse 82, CH-4056 Basel, Switzerland}

\author{Sebasti\'{a}n A. D\'{i}az}
\affiliation{Department of Physics, University of Basel, Klingelbergstrasse 82, CH-4056 Basel, Switzerland}

\author{Jelena Klinovaja}
\affiliation{Department of Physics, University of Basel, Klingelbergstrasse 82, CH-4056 Basel, Switzerland}

\author{Daniel Loss}
\affiliation{Department of Physics, University of Basel, Klingelbergstrasse 82, CH-4056 Basel, Switzerland}

\begin{abstract}
	When interacting spins in condensed matter order ferromagnetically, their ground state wave function is topologically trivial. Nonetheless, in two dimensions, the ferromagnetic state can support spin excitations with nontrivial topology, an exotic state known as topological magnon insulator (TMI). 
	Here, we theoretically unveil and numerically confirm a novel ferromagnetic state in three dimensions dubbed second-order TMI, whose hallmarks are excitations at its hinges, where facets intersect. 
	Since ferromagnetism naturally comes with broken time-reversal symmetry, the hinge magnons are chiral, rendering backscattering impossible. Hence, they trace out a three-dimensional path about the sample unimpeded by defects and are topologically protected by the spectral gap. They are remarkably robust against disorder and simultaneously highly tunable by atomic-level engineering of the sample termination.
	Our findings empower magnonics with the tools of higher-order topology, a promising route to combine low-energy information transfer free of Joule heating with three-dimensional vertical integration. 
\end{abstract}

\maketitle

The quantum Hall effect and the Chern-insulating state of electrons are two of the great discoveries in the second half of the 20th century that have shaped today's solid state research by amalgating Bloch's band theory with quantum state geometry and topology\cite{Klitzing1986, pankratov1987supersymmetric, Haldane1988}.
One of the many novel exotic phases of matter brought to light by this fruitful synthesis is that of the topological magnon insulator (TMI), a two-dimensional phase that exhibits a spectrum of topologically nontrivial bosonic excitations, called magnons, above a topologically trivial magnetically ordered ground state \cite{Zhang2013, Shindou13, Mook14b,Molina2016, Nakata2017QSHE, Diaz2019}. A topologically nontrivial gap in the magnon spectrum protects magnonic edge states. Due to time-reversal violation in ferromagnets, the propagation of the edge modes is chiral akin to electronic Chern insulators, giving rise to magnon Hall effects \cite{Meier2003, Katsura2010, Matsumoto2011a, Hoogdalem2013, Mook14a, Nakata2016, Mook2018, Kim2019}. Hence, TMIs support unidirectional magnon currents that---once coherently excited---transfer information along the sample's boundary \cite{Zhang2013, Shindou13}. In sharp contrast to electrons, the charge neutral magnonic currents do not cause Ohmic heating \cite{Chumak2015}, promising low-energy information transfer and giving rise to the paradigm of ``topological magnonics'' \cite{Wang2018TopMagnonics, Yamamoto2019, Diaz2020, Aguilera2020}.
However, since the magnonic Chern insulator is a two-dimensional phase of matter, it is not suitable to keep up with CMOS electronics design trends such as three-dimensional vertical integration \cite{Topol2006}.

\begin{figure}
        \centering
        \includegraphics[scale = 1]{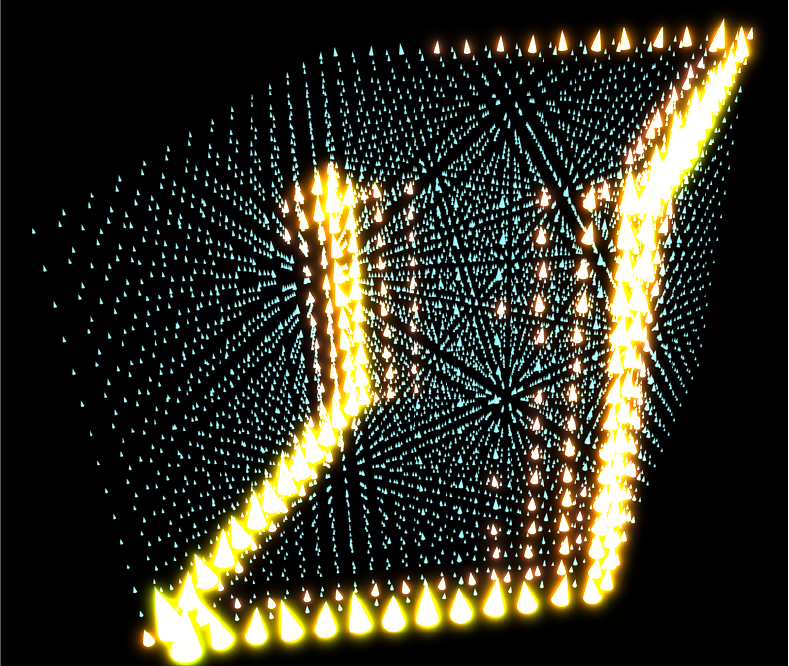}
		\caption{Snapshot of a chiral hinge magnon in a SOTMI visualized by atomistic spin dynamics simulations. At each lattice site of a stack of honeycomb layers, a classical spin vector is represented by a little cone, whose size encodes its deviation from the ferromagnetic ground state. Large cones indicate strongly excited spins. Since the magnon spectrum exhibits a gap, within which topologically protected states only exist at the hinges of the sample, a coherent local excitation at one of the hinges (here: in the middle of the rearward left hinge) launches a unidirectionally propagating spin wave. The snapshot is taken before the spin wave completed the loop along the hinges. 
		The topological protection due to the absence of backscattering renders the chiral hinge magnon remarkably robust against defects and disorder.
    }
    	\label{fig:artistic}
\end{figure}

Herein, we contribute to the foundations of topological magnonics by reporting our theoretical discovery of a novel exotic phase of matter in three dimensions dubbed second-order TMI (SOTMI). In general, the hallmark of a higher-order (or $n$th-order) topological phase in $d$ dimensions are gapless states at its $n$th-order boundaries ($n \ge 2$) \cite{Benalcazar2017, Schindler2018}. So far, second-order ($n=2$) topological magnons have been identified as corner states in two-dimensional magnets\cite{Li2019HOTISoliton, Sil2020, Hirosawa2020TopQuadrupole}. In contrast, we present a SOTMI in three dimensions, whose hinges, the intersections of facets, support gapless chiral magnons, as depicted in Fig.~\ref{fig:artistic}. These hinge magnons trace out a three-dimensional path, allowing for magnonic information transfer in all spatial directions. We explicitly simulate hinge magnons in the presence of disorder and perturbations that break crystalline symmetries, unveiling their remarkable topological robustness owed to their chirality.
Nonetheless, their path in real space turns out to be highly tunable by a manipulation of the surface termination at the atomic level. Thus, our findings empower magnonics with the tools of higher-order topology.

\begin{figure}[t]
    \centering
    \includegraphics[scale = 1]{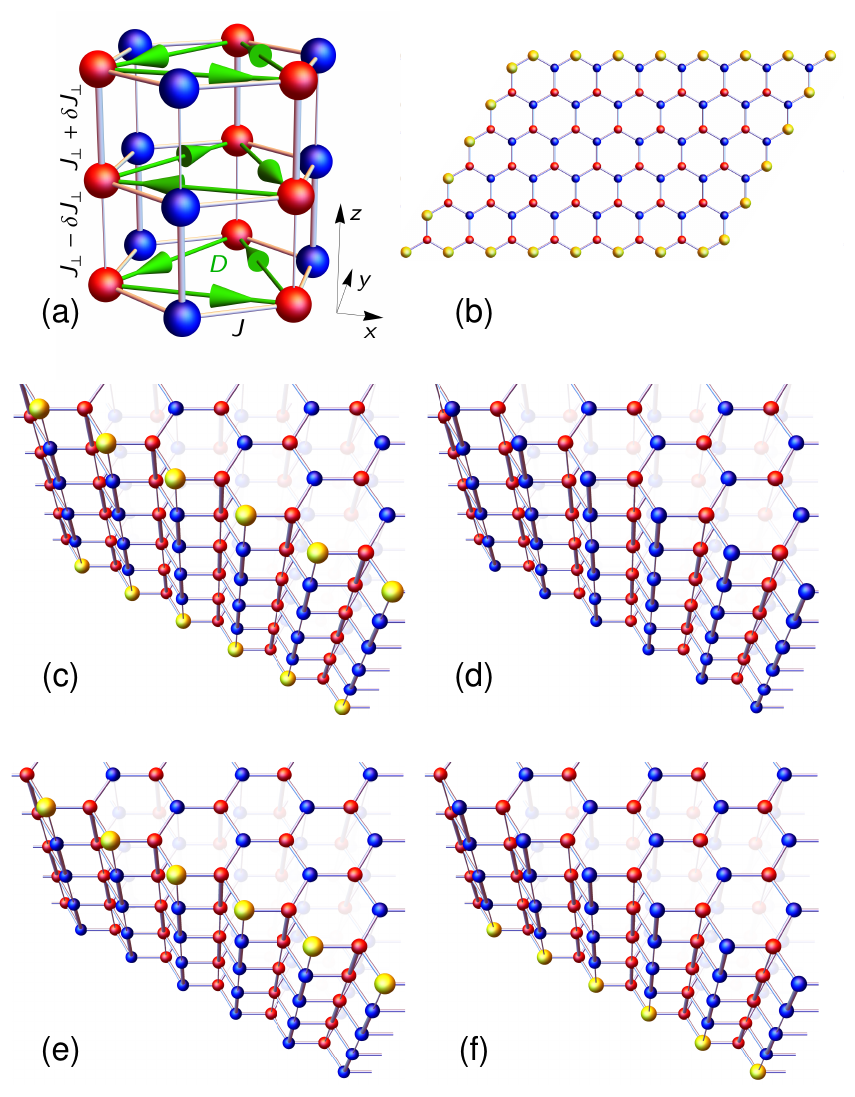}
    \caption{Microscopic model of a SOTMI with chiral hinge states in three dimensions. 
    (a) A stack of honeycomb layers with indicated magnetic interactions. Blue and red spheres indicate the A and B sublattice of the honeycomb, respectively.
    (b) A single honeycomb layer realizes a magnon Chern insulator, whose hallmark is a gap in the magnon spectrum bridged only by a chiral edge state (yellow spheres).
    (c-f) View along a zigzag-terminated surface of stacks built from a finite number of layers. Even-numbered stacks exhibit either (c) hinge modes at both terminating layers  or (d) no hinge modes at all.
    In contrast, odd-layered stacks exhibit hinge states at one of the terminating layers, either at the top (e) or the bottom (f) of the stack.}
    \label{fig:SSH}
\end{figure}

We consider a stack of honeycomb magnets with spins situated at the honeycomb's vertices, as indicated by spheres in Fig.~\ref{fig:SSH}(a). The interactions between these spins are comprised in the Hamiltonian 
\begin{align}
	H  = H_\mathrm{\parallel} + H_\mathrm{\perp} + H_\mathrm{\perp}^\mathrm{\delta} + H_\text{Z}.
	\label{eq:hamiltonian}
\end{align}
Here, intralayer interactions $H_\mathrm{\parallel} = \sum_l h_\mathrm{\parallel}^{(l)}$ ($l$ is the layer index), with
\begin{align}
    h_\mathrm{\parallel}^{(l)}
    = 
    - \frac{J}{2} \sum_{\langle ij \rangle} \boldsymbol{S}_{i}^{(l)} \cdot \boldsymbol{S}_{j}^{(l)} 
    + \frac{(-1)^l D}{2} \sum_{\langle \langle ij \rangle \rangle} \nu_{ij} \hat{\boldsymbol{z}} \cdot \boldsymbol{S}_{i}^{(l)} \times \boldsymbol{S}_{j}^{(l)},
\end{align}
include positive nearest neighbor exchange interaction $J$ that stabilizes ferromagnetic order. 
Upon a magnon expansion (see Methods) a single layer is found to feature two magnon branches, resembling the graphene band structure with Dirac cones. The latter acquire a topological mass gap~\cite{Owerre2016a} $\pm 6\sqrt{3} D/J$ (later referred to as ``bulk gap'') by next-nearest neighbor Dzyaloshinskii-Moriya interaction\cite{Dzyaloshinsky58, Moriya60} (DMI) $D$ [green arrows in Fig.~\ref{fig:SSH}(a)]; $\hat{\boldsymbol z}$ is a unit vector along the $z$ direction and $\nu_{ij} = \pm 1$, with $+$ ($-$) for counterclockwise (clockwise) circulation.
The topological nontriviality is captured by a nonzero winding number $w^{(l)} = (-1)^l \mathrm{sgn}( m D )$ (see Methods). Here, $m=+1$ ($m=-1$) for the ferromagnetic ground state pointing along the positive (negative) $z$ direction.
Hence, a single layer exhibits chiral edge states, as indicated by yellow spheres in Fig.~\ref{fig:SSH}(b). From hereinafter, we consider $m=+1$.

\begin{figure}[t!]
    \centering
    \includegraphics[scale=1]{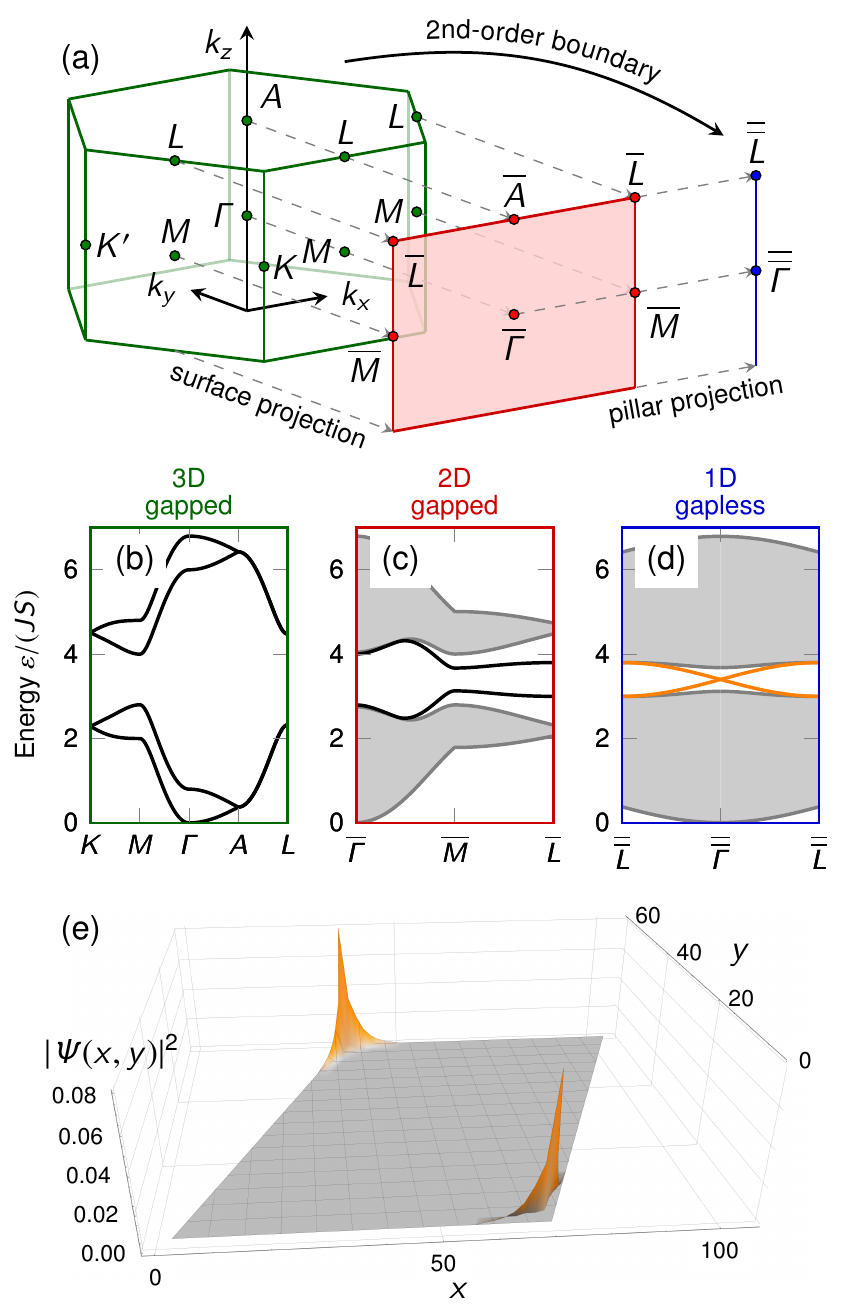}
    \caption{
    Brillouin zones (BZs) and magnon band structures in 3D, 2D, and 1D.
    (a) For a successive reduction of dimensions, the hexagonal 3D BZ (green) first gets projected onto a surface to yield the 2D BZ (red), which then is projected onto a line, resulting in a 1D BZ (blue). Selected high-symmetry points are indicated. 
    (b) Gapped 3D spectrum showing four magnon branches. (c) Gapped 2D spectrum, with gray areas indicating the bulk continuum projected onto the $xz$ surface and black lines indicating surface states. (d) Gapless 1D spectrum with chiral hinge magnons crossing the band gap (orange lines). Gray areas indicate both the bulk and surface continuum projected onto the $z$ axis.
    (e) Probability density $|\varPsi(x,y)|^2$ of the chiral hinge-magnon states in real space for a pillar with a parallelogram cross section [cf.~Fig.~\ref{fig:SSH}(b)] at $\overline{\overline{\varGamma}}$. Open boundary conditions are assumed in both $x$ and $y$ direction but periodic boundary conditions in $z$ direction. The two chiral modes are localized at opposite obtuse corners of the pillar. 
    Parameters read $d = j_\perp = \delta j_\perp = 0.2$ and $b=0$.
    }
    \label{fig:bz-and-bands}
\end{figure}

The AA-stacked honeycomb layers are coupled by
\begin{align}
	H_\perp
    = 
    -J_\perp \sum_i \sum_{l} \boldsymbol{S}_{i}^{(l)} \cdot \boldsymbol{S}_{i}^{(l+1)},
\end{align}
with ferromagnetic interlayer exchange $J_\perp$ due to which magnons obtain a dispersion along the stacking direction. We assume that $J_\perp$ is sufficiently small such that the bulk gap due to $D$ stays open, which is a reasonable assumption for layered structures \cite{Zhang2019vanderWaalsReview}. Using the Brillouin zone convention in Fig.~\ref{fig:bz-and-bands}(a), a representative bulk magnon spectrum is shown in Fig.~\ref{fig:bz-and-bands}(b); notice the band gap between the lower and upper pair of bands. 

Letting the sign of DMI alternate between adjacent layers ensures alternating winding numbers and chiral edge states, which gap out pairwise. However, since each layer is a mirror plane for the infinite stack, a magnonic surface Dirac cone is stabilized, rendering the surface spectrum gapless. Hence, so far, the stack is a (first-order) topological mirror insulator \cite{Teo2008, Fulga2016} of magnons [see Supplementary Information (SI) \cite{Supplement}].
To break this mirror symmetry, we effectively buckle each layer. This is accounted for by an alternating modulation of the interlayer exchange interaction ($\delta J_\perp$),
\begin{align}
    H_\perp^\delta
    = 
    \delta J_\perp \sum_{l} (-1)^l \left( \sum_{i \in \mathrm{A}} \boldsymbol{S}_i^{(l)} \cdot \boldsymbol{S}_i^{(l+1)} 
    - \sum_{i \in \mathrm{B}} \boldsymbol{S}_i^{(l)} \cdot \boldsymbol{S}_i^{(l+1)} \right),
\end{align}
which is opposite for the A and B sublattice of the honeycomb [cf.~alternating interlayer bonds between blue and red sites in Fig.~\ref{fig:SSH}(a)].
Finally, a magnetic field $B^z<0$ ($m=+1$) is applied that enters the Zeeman Hamiltonian 
\begin{align}
    H_\text{Z} = \sum_l \sum_i B^z S_{i}^{(l), z}.
\end{align}
Using reduced constants $b = B^z/(JS)$, $j_\perp = J_\perp/J$, $d=D/J$, and $\delta j_\perp = \delta J_\perp / J$, the ferromagnetic ground state is stable for $|\delta j_\perp| \le \frac{1}{2} \sqrt{(-b+2j_\perp)(-b+2j_\perp +6)}$; note that $b \le 0$. Below, we consider parameters $d = j_\perp = \delta j_\perp = 0.2$, for which the ferromagnetic state is stable even at $b=0$; see SI \cite{Supplement} for other cases.

The mirror-symmetry breaking $\delta j_\perp$ acts like a mass that gaps out surface states, as depicted in Fig.~\ref{fig:bz-and-bands}(c). This can be understood in the following intuitive way. It is well-established that the wave function of edge states in graphene has weight predominantly on the sublattice whose atoms dominate in a particular termination \cite{CastroNeto2009}. For example, the edge states of zigzag-terminated graphene live on that sublattice whose atoms constitute the very edge. Now, consider a stack of an even number of layers with a zigzag-terminated surface, as shown in Fig.~\ref{fig:SSH}(c). The surface state has weight mainly on the blue sublattice. Along the stacking direction, spins located at the blue sites of the chains resemble a spin version of the Su-Schrieffer-Heeger (SSH) model \cite{Su1979}. The sign of $\delta j_\perp$ determines whether the chain is topologically trivial or nontrivial, i.e., if its ends feature bound states [Fig.~\ref{fig:SSH}(c), $\delta j_\perp > 0$] or not [Fig.~\ref{fig:SSH}(d), $\delta j_\perp < 0$]. 
For an odd number of honeycomb layers, there is always one undimerized dangling spin hosting the bound state, either at the top [Fig.~\ref{fig:SSH}(e), $\delta j_\perp > 0$] or bottom layer [Fig.~\ref{fig:SSH}(f), $\delta j_\perp < 0$]. 
Due to the intralayer coupling of SSH-like chains, the states bound to the chains' ends can propagate along the hinge to which they are confined. The DMI-induced chirality of each layer admits propagation only in one direction, promoting the end states to chiral hinge modes.

A different surface of a finite stack may be terminated by the red sublattice that exhibits the opposite dimerization pattern. Hence, whatever sign of $\delta j_\perp$ causes bound states at a particular end of the blue chains, leads to the red chains not hosting bound states at this very end (and vice versa).
If domains of opposite surface terminations (of ``opposite color'') meet, there is a domain wall between a topologically trivial and nontrivial phase, necessitating a gapless mode along the domain wall, i.e., along the stacking direction. Such domain walls naturally occur at the hinges of materials where facets intersect, suggesting the name ``chiral hinge magnons''. In the spectrum of an infinite pillar (i.e., an infinite stack of honeycomb layers of finite size), hinge magnons appear as bands that connect adjacent bulk and surface bands, as shown in Fig.~\ref{fig:bz-and-bands}(d). The 
hinge-magnon wave function is strongly localized to the hinges [cf.~Fig.~\ref{fig:bz-and-bands}(e)].

\begin{figure*}
    \centering
	\includegraphics[width = 0.8\textwidth]{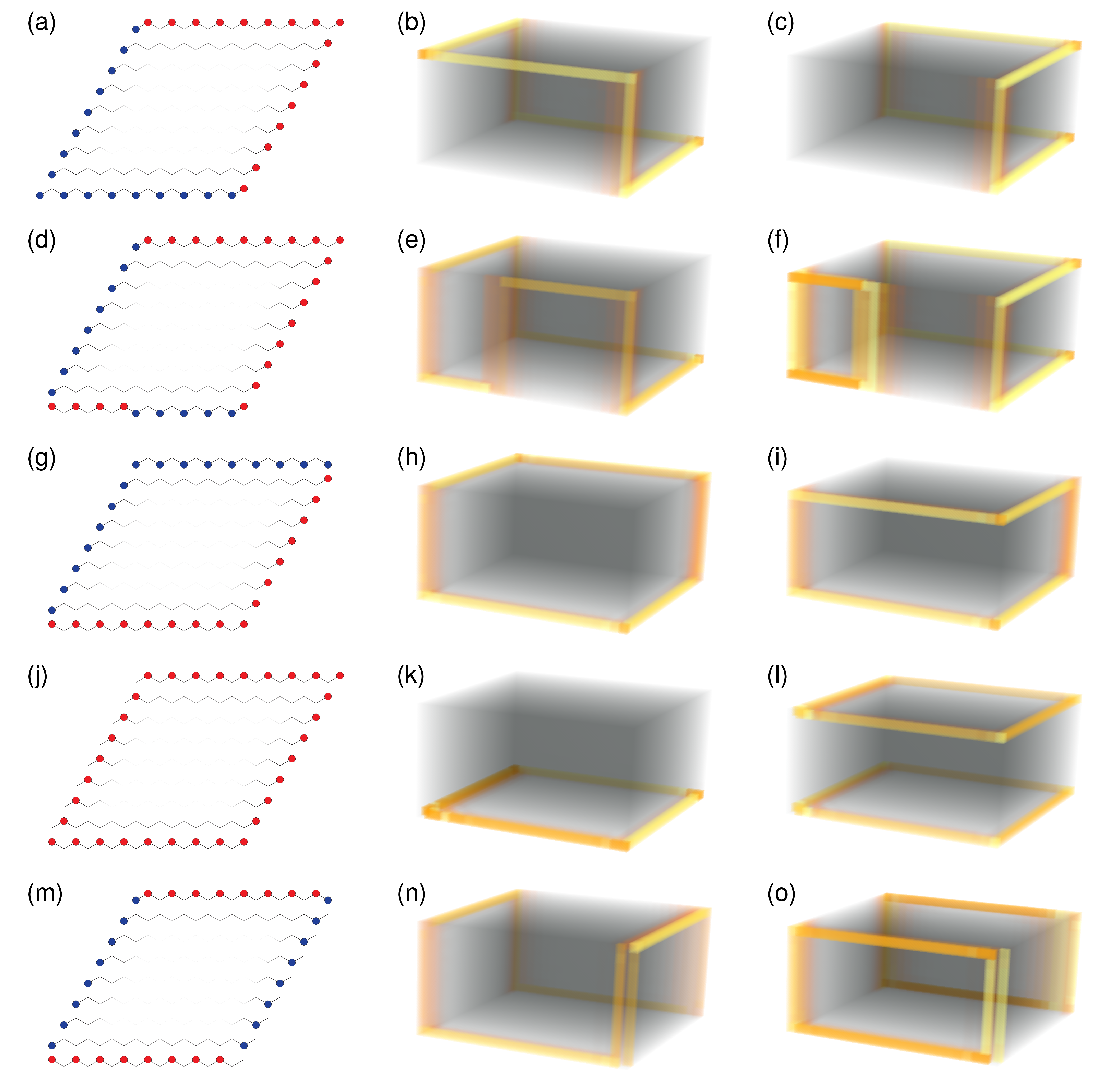}
	\caption{
	Chiral hinge magnons in finite-sized stacks of honeycomb-lattice ferromagnets for selected terminations.
	(Left column) Sketch of stacks' cross sections with blue and red circles indicating the termination spanning the full height of the stack. (Central/right column) Probability density of hinge magnons in a finite stack built from an odd (13)/even (12) number of layers. Each layer consists of $25 \times 25$ honeycomb unit cells. Black transparent/orange opaque color indicates zero/maximal probability density of the hinge magnon. The view angle is chosen such that the frontmost hinge coincides with the lower right corner of the cross sections. (a,b,c) All-zigzag terminated cross section with domain walls at the obtuse corners of the parallelogram. (d,e,f) Cross section with several blue vertical chains removed, giving rise to a new termination domain and, hence, two new domain walls. (g,h,i) Upon removing all terminating sites at two opposite boundaries, the domain walls get shifted to the acute corners. (j,k,l) Removing only the blue sites [compared to (a)] causes a uniform termination without domain walls. (m,n,o) Removing the terminating sites from the surfaces enclosing the frontmost obtuse corner [compared to (a)] causes domain walls at all hinges. Parameters read $d = j_\perp = \delta j_\perp = 0.2$ and $b=0$.}
	\label{fig:HingeModes}
\end{figure*}

We present selected examples in Fig.~\ref{fig:HingeModes}. For a stack with a parallelogram cross-section and all-zigzag termination, as shown in Fig.~\ref{fig:HingeModes}(a), the termination changes from blue to red at the obtuse corners. The associated hinges feature a domain wall and, hence, also a chiral hinge magnon, a prediction that is confirmed numerically by exact diagonalization of a finite sample by means of linear spin-wave theory [see Fig.~\ref{fig:HingeModes}(b,c)]. While the position of the hinge magnons is tied to the domain walls, it is the number of layers that determines the actual path taken. This is because a stack of an odd number of layers has a nonzero net winding number, originating from one layer being uncompensated \cite{Takahashi2020}. Hence, there must be one chiral mode circulating the stack about the stacking direction [cf.~Fig.~\ref{fig:HingeModes}(b)]. In contrast, an even number of layers has a net winding number of zero, ruling out any net chirality about the stacking direction [cf.~Fig.~\ref{fig:HingeModes}(c)]. Nonetheless, the hinge magnons are still chiral as their propagation is unidirectional.

Upon removing a couple of vertical blue chains from a surface, a red domain arises within the formerly blue-terminated surfaces [Fig.~\ref{fig:HingeModes}(d)]. Two new mass domain walls are created, forcing the hinge magnon to take a detour in stacks with an odd number of layers [Fig.~\ref{fig:HingeModes}(e)]. For an even number of layers, two independent chiral hinge magnons, amounting to two separate loops, are found [Fig.~\ref{fig:HingeModes}(f)].

Removing all terminating spins from two opposite surfaces, as depicted in Fig.~\ref{fig:HingeModes}(g), results in the domain walls being shifted to the acute corners. The hinge magnons redistribute accordingly [see Figs.~\ref{fig:HingeModes}(h,i)]. Hence, termination manipulations at the atomic level allow to engineer samples with hinge modes at arbitrary hinges. In particular, one may remove any domain walls [Fig.~\ref{fig:HingeModes}(j)], resulting in the chiral modes not crossing the stack at all [Figs.~\ref{fig:HingeModes}(k,l)]. Similarly, domain walls at all hinges [Fig.~\ref{fig:HingeModes}(m)] cause chiral hinge magnons at all hinges [Figs.~\ref{fig:HingeModes}(n,o)]. We reiterate that the path of the hinge magnons depends also on the sign of $\delta j_\perp$, as we explicitly show in the SI \cite{Supplement}.

For the very special case that the Hamiltonian respects inversion symmetry, the existence of chiral hinge magnons is captured by the recently developed machinery of higher-order topology\cite{Benalcazar2017,  Schindler2018}, which associates a bulk topological number with the hinge modes. We show in the Methods how to apply these tools to the present magnonic case but point out that inversion symmetry is not a prerequisite for chiral hinge magnons. As a matter of fact, we show in the SI that the chiral hinge magnons are remarkably robust against various types of inversion-symmetry-breaking bulk spin interactions \cite{Supplement}.

\begin{figure}
    \centering
    \includegraphics[scale=1]{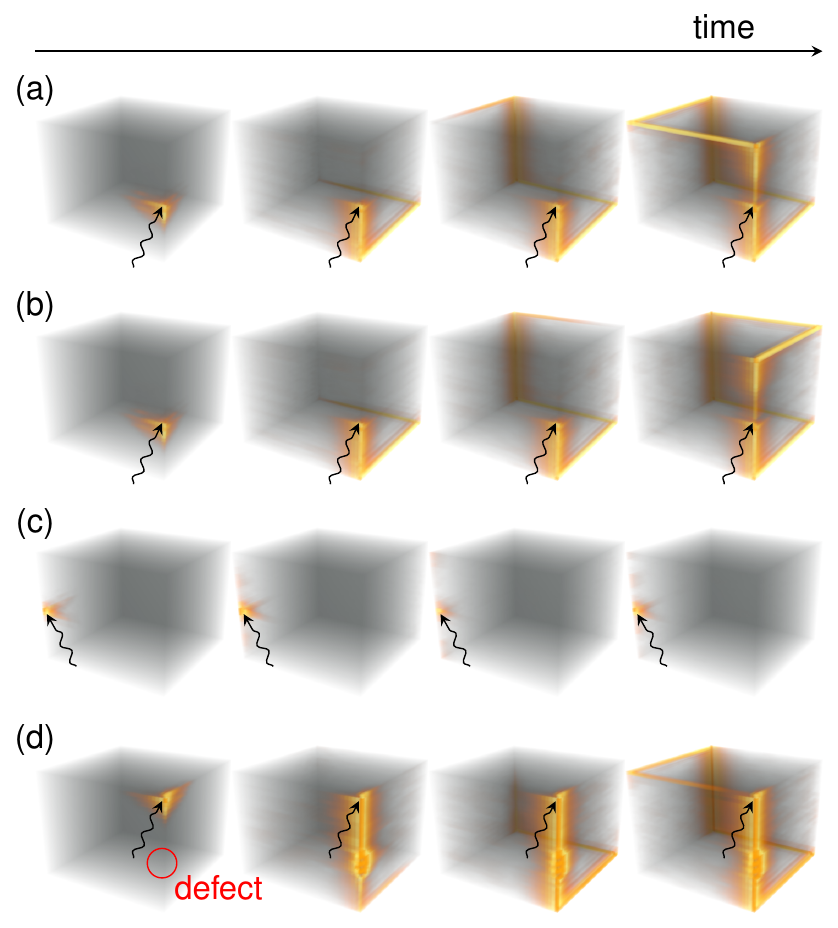}
    \caption{
    	Numerical simulation of a finite-sized SOTMI built from (a,c,d) $31$ or (b) $30$ layers. There are $40\times 40$
	honeycomb unit cells per layer. Snapshots show the time evolution upon a local coherent excitation (indicated by the wavy arrow) with an energy within the global band gap of the magnon spectrum [cf.~Fig.~\ref{fig:bz-and-bands}(d)]. Black transparent/orange opaque color indicates zero/maximal probability density.
    	(a) Upon exciting a single spin at a domain wall a topologically protected chiral magnon propagates around the sample along the hinges.
    	(b) For an even number of layers, a similar excitation causes the hinge magnon to take a different path around the sample.
    	(c) Excitations at hinges without a domain wall merely cause evanescent waves.
    	(d) Due to the impossibility of backscattering the chiral excitation bypasses defects at the hinges.
    	For movies, see (a) \texttt{1-obtuse.mp4}, (b) \texttt{2-even.mp4}, (c) \texttt{3-acute.mp4}, and (d) \texttt{4-defect.mp4} in the SI\cite{Supplement}.
    	The simulations are based on the Landau-Lifshitz equation without damping. Parameters read $d = j_\perp = \delta j_\perp = 0.2$ and $b=0$.
    }
    \label{fig:dynamics}
\end{figure}

The above analysis relied on the magnon wave function as obtained within linear spin-wave theory. Next, we present independent numerical evidence for chiral hinge magnons by simulating a coherent excitation experiment by means of atomistic spin dynamics simulations (see Methods). We consider a stack similar to that in Figs.~\ref{fig:HingeModes}(a,b,c).
An ac magnetic field with a frequency $\varepsilon_\text{ex} = 3.4JS$ within the global band gap [where the hinge magnons cross the $\overline{\overline{\varGamma}}$ point in Fig.~\ref{fig:bz-and-bands}(d)] applied to a single spin at the obtuse hinges excites the hinge magnon. Its chiral information transfer along the three-dimensional path can be clearly traced [Fig.~\ref{fig:dynamics}(a,b)].  
In contrast, a local excitation at the acute hinges does not result in chiral information transfer, but rather in an evanescent wave [Fig.~\ref{fig:dynamics}(c)]. This finding complies with the absence of probability density of the chiral in-gap states at the acute corners [cf.~Figs.~\ref{fig:bz-and-bands}(e) and \ref{fig:HingeModes}(b,c)]. The simulations also reveal the hinge magnon's robustness against backscattering at defects [Fig.~\ref{fig:dynamics}(d)]. For the hinge magnon to scatter into states with opposite momentum, it would have to scatter to the opposite hinge, a process that is exponentially suppressed by spatial separation. Hence, chiral magnon information transfer is immune to defects.

\begin{figure*}[t]
    \centering
    \includegraphics[width=0.8\textwidth]{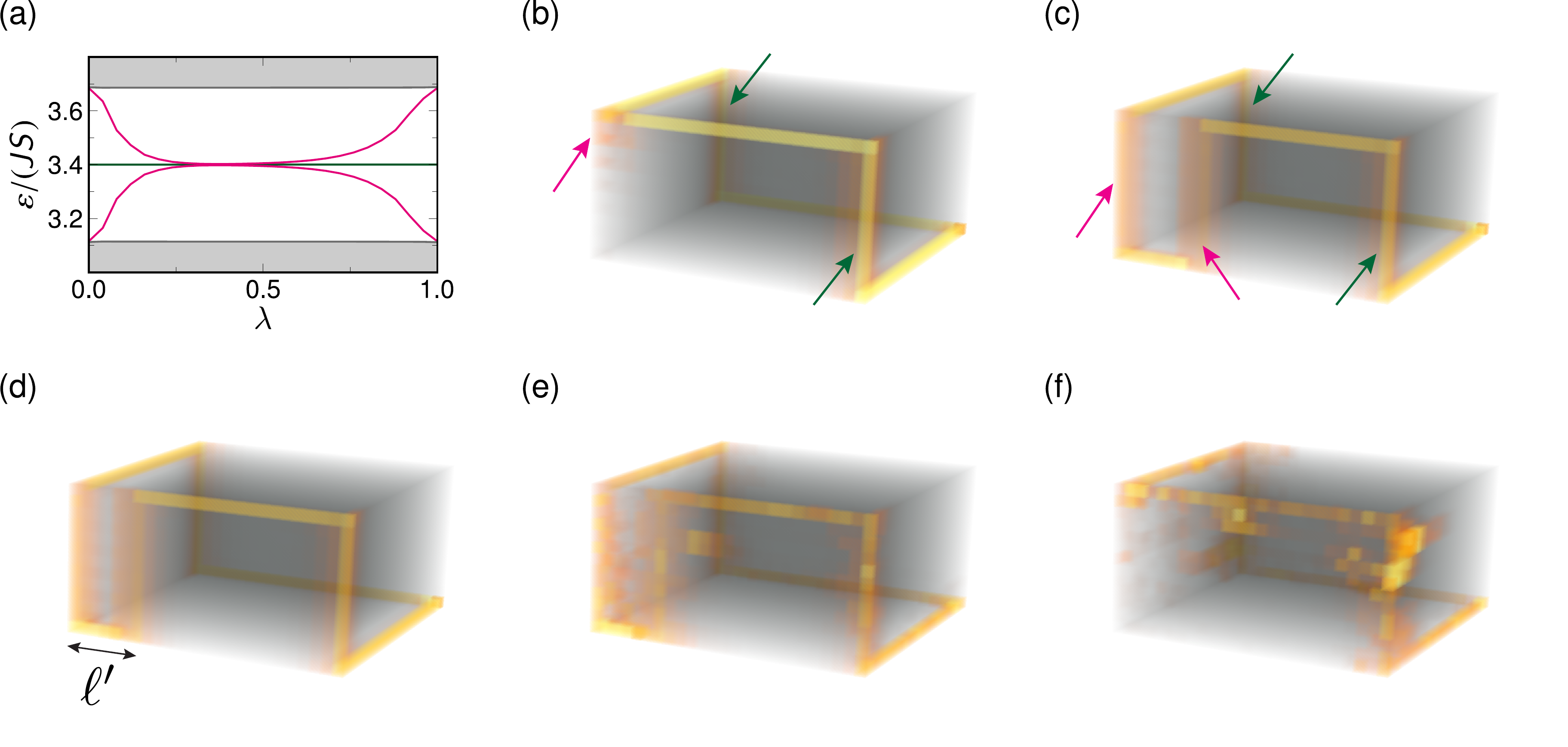}
    \caption{Finite-size and disorder effects on chiral hinge magnons in SOTMIs. 
    (a) Gapless 1D spectrum of an infinite stack at the $\overline{\overline{\varGamma}}$ point in dependence on $\lambda$ which parametrizes the process of removing terminating vertical chains from one surface, a process similar to Fig.~\ref{fig:HingeModes}(d). $\lambda = 0$: no chains removed [blue termination in Fig.~\ref{fig:HingeModes}(d)]. $\lambda = 1$: all chains removed (red termination). Spatially well-separated hinge magnons appear as a green horizontal line. Two new hinge states associated with the new domain wall are colored in magenta. The gray continuum indicates projected surface states. 
    (b,c) Probability density of the hinge magnon in a finite stack of an odd number (13) of layers for (b) $\lambda \approx 0.12$ and (c) $\lambda \approx 0.32$.
    (d-f) Probability density of the hinge magnon for $\lambda \approx 0.24$ and increasing disorder; (d) $\sigma_b/\Delta = 0$; (e) $\sigma_b/\Delta = 1.11$; (f) $\sigma_b/\Delta = 1.48$. The hinge modes due to short domains ($\ell'$) gap out for sufficiently large disorder.
    Parameters read $d = j_\perp = \delta j_\perp = 0.2$ and $b=0$ and the pillar's cross section is built from $25\times 25$  honeycomb unit cells per layer.
     }
    \label{fig:disorder}
\end{figure*}

The hinge magnons' localization to domain walls may be quantified by a localization length $\xi$, which is inversely proportional to the surface gap $\Delta \propto \delta j_\perp$ from broken mirror symmetry, thus, $\xi \propto 1/\delta j_\perp$. For large enough domains, of size $\ell \gg \xi$, neighboring counter-propagating hinge magnons are well-separated and do not hybridize. However, if the boundary consists of domains $\ell \approx \xi$, hinge states of opposite chirality overlap and gap out. For example, consider the situation in Fig.~\ref{fig:HingeModes}(d) as a gradual process parametrized by $\lambda \in [0,1]$. Starting with no vertical chains removed ($\lambda = 0$), one chain at a time is removed, until the termination of the manipulated surface has fully changed from zigzag (blue) to bearded (red) ($\lambda = 1$). Figure \ref{fig:disorder}(a) shows the magnon spectrum of an infinite stack at the $\overline{\overline{\varGamma}}$ point [where the hinge modes cross, cf.~Fig.~\ref{fig:bz-and-bands}(d)] in dependence on $\lambda$. Two degenerate states are found at $3.4/(JS)$ (green line) corresponding to well separated hinge modes [green arrows in Fig.~\ref{fig:disorder}(b)]. The surface gap hosts two states that split off for $\lambda$ close to $0$ or $1$ [magenta lines in Fig.~\ref{fig:disorder}(a)]. In these limits, two domain walls come close together, causing their hinge magnons to overlap and gap out [magenta arrow in Fig.~\ref{fig:disorder}(b)]. However, around $\lambda =0.5$, the domain sizes are sufficiently large to suppress finite size effects and to enforce the chiral magnon to take a detour, as depicted in Fig.~\ref{fig:disorder}(c).
Thus, it is $\xi$ (or $\Delta$) what sets the lower threshold for miniaturization of devices to support chiral hinge magnons.

The size of the gap $\Delta$ also protects the chiral hinge magnons against disorder, whose strength we denote by $\sigma_b$ (see Methods). 
For example, Fig.~\ref{fig:disorder}(d) shows the probability density of a hinge magnon in the absence of disorder. The hinge magnon takes a detour around a small domain wall of length $\ell' > \xi$. As disorder is increasing, the effective, disorder-averaged gap $\Delta(\sigma_b)$ is decreasing, leading to an increasing $\xi(\sigma_b)$ [see Fig.~\ref{fig:disorder}(e)]. Once $\xi(\sigma_b) \gtrsim \ell'$ the associated hinge magnons gap out [see Fig.~\ref{fig:disorder}(f)].
Consequently, larger domain walls are more robust against disorder than smaller domains. Eventually, very strong disorder closes the gap and causes localization. The effects of disorder are also captured by spin dynamics simulations, presented in the SI \cite{Supplement} (for movies, see \texttt{5-disorder-$R$.mp4}).

With the influence of disorder and defects suppressed by the topological gap, only a finite magnon lifetime $\tau = \hbar /(2 \alpha \varepsilon)$ due to ubiquitous intrinsic Gilbert damping $\alpha$ remains, originating from phononic or, in metals, electronic baths \cite{Brataas2008}. For typical values $S=3/2$, $J \approx \unit[2.2]{meV}$, and $D/J \approx 0.1$ (resembling those obtained for CrI$_3$\cite{Chen2020}), the hinge-magnon energy is approximated by $\varepsilon \approx 3 S J$, resulting in $\tau \approx \unit[330]{ps}$ for $\alpha = 10^{-4}$. With a velocity up to $v = \unit[500]{m/s}$, a mean free path up to $v \tau \approx \unit[165]{nm}$ is obtained. As long as the operating temperature is well below the ordering temperature, the effects of temperature are negligible (see Methods). Hence, a realization of the SOTMI state by relying on recent advances in atomic-scale magnonic crystals \cite{Qin2019Zakeri}, in magnetic silicene \cite{Tokmachev2018}, in van der Waals magnets\cite{Zhang2019vanderWaalsReview} such as CrI$_3$ \cite{Huang2017} or magnetic organic materials \cite{Liu2018orgmag}, may find application in nano-scale exchange magnonics.
One may also abandon the atomic scale and implement SOTMIs in magnonic metamaterials built from three-dimensional coupled arrays of spin torque oscillators \cite{Talatchian2020}, magnetic vortex structures \cite{Pulecio2014}, magnonic quantum networks \cite{Rusconi2019}, or superconducting spin qubits \cite{Cai2019}. In particular, one may stack two-dimensional topological magnonic crystals realized either as iron islands in an yttrium iron garnet matrix \cite{Shindou13} or as a patterned ferrimagnetic insulator \cite{Li2018magcrys}. The alternating sign of the Chern number is arranged for by varying the aspect ratio of the islands and rotating the patterns, respectively. Gigahertz hinge magnons with mean free paths up to several millimeters are expected. 

We proposed a novel topological phase of matter dubbed SOTMI to be realized either in magnetic metals or insulators. Its hallmarks are chiral magnon states along its hinges, a finding that opens up possibilities to design innately 3D ``information highways,'' circumventing the problem of nontopological magnon propagation in bent magnonic waveguides \cite{Vogt2012}. Thus, besides being an exciting and exotic second-order topological phase that could allow to experimentally test the foundations of higher-order topology, SOTMIs add to the arsenal of 3D magnonics \cite{Gubbiotti2019} to compete with today's CMOS technology design trends such as vertical integration.

\begin{acknowledgments}
This work was supported by the Georg H.~Endress Foundation, the Swiss National Science Foundation, and NCCR QSIT. This project received funding from the European Unions Horizon 2020 research and innovation program (ERC Starting Grant, Grant No 757725).
\end{acknowledgments}

\section*{Methods}
\appendix

\section{Linear spin-wave theory}
The excitations above a magnetically ordered ground state can be addressed within spin-wave theory, whose main idea is to map the spin operators $\boldsymbol{S}_i$ onto bosonic creation and annihilation operators $a_i^\dagger$ and $a_i$. In the limit of low temperatures, within which the density of excitations is sufficiently small that interactions between them may be neglected, a truncated Holstein-Primakoff transformation \cite{Holstein1940}
\begin{subequations}
\begin{align}
    S_i^x &\approx \sqrt{\frac{S}{2}} \left( a_i + a_i^\dagger \right),
    \\
    S_i^y &\approx - \mathrm{i} \sqrt{\frac{S}{2}} \left( a_i - a_i^\dagger \right),
    \\
    S_i^z &= S - a_i^\dagger a_i,
\end{align}
\end{subequations}
is appropriate; $\mathrm{i}^2 = -1$. For the model under consideration, there are four spins in the basis because the dimerization pattern due to $\delta J_\perp$ doubles the honeycomb unit cell in the stacking direction. The bosonic operators may be labelled $a_{n,\boldsymbol{R}_i}$, where $\boldsymbol{R}_i$ is the coordinate vector of the $i$th magnetic basis and $n=1,\ldots,4$ enumerates the $n$th basis spin.
After a Fourier transformation
\begin{align}
    a_{n,\boldsymbol{R}_i} = \frac{1}{\sqrt{N}} \sum_{i=1}^N \mathrm{e}^{\mathrm{i} \boldsymbol{k} \cdot \boldsymbol{R}_i} a_{n,\boldsymbol{k}}, 
\end{align}
to momentum $\boldsymbol{k}$, where $N$ is the number of unit cells, the Hamiltonian reads
$
    H - E_0 
    \approx 
    H_2 =
    \sum_{\boldsymbol{k}} \boldsymbol{\varPsi}_{\boldsymbol{k}}^\dagger \cdot \mathsf{H}_{\boldsymbol{k}} \cdot \boldsymbol{\varPsi}_{\boldsymbol{k}}
$.
Here, $E_0$ is the unimportant classical ground state energy and 
$
    \boldsymbol{\varPsi}_{\boldsymbol{k}}^\mathrm{T} = (a_{1,\boldsymbol{k}}, a_{2,\boldsymbol{k}}, a_{3,\boldsymbol{k}}, a_{4,\boldsymbol{k}})
$
a vector built from Holstein-Primakoff bosons associated with the four sublattices. The Fourier kernel of the bilinear Hamiltonian $H_2$ reads
\begin{align}
    \mathsf{H}_{\boldsymbol{k}} 
    = 
    S J
    \begin{pmatrix}
        \mathsf{H}_{\boldsymbol{k}}^{\parallel,+} & \mathsf{H}_{\boldsymbol{k}}^\perp \\
        (\mathsf{H}_{\boldsymbol{k}}^\perp)^\ast & \mathsf{H}_{\boldsymbol{k}}^{\parallel,-}
    \end{pmatrix},
    \label{eq:ham-kernel}
\end{align}
with its intralayer, $\mathsf{H}_{\boldsymbol{k}}^{\parallel,\pm}$ and interlayer submatrices, $\mathsf{H}_{\boldsymbol{k}}^\perp$, given in the SI \cite{Supplement}. Upon diagonalization of $\mathsf{H}_{\boldsymbol{k}}$, one obtains the magnon energies $\varepsilon_{n,\boldsymbol{k}}$.

\section{Winding number of a single honeycomb layer}
After a Holstein-Primakoff expansion about the ferromagnetic state polarized along the $z$ direction, the bilinear magnon Hamiltonian of a single honeycomb layer reads 
$
	H_2 =
    \sum_{\boldsymbol{k}} \boldsymbol{\varPhi}_{\boldsymbol{k}}^\dagger \cdot \tilde{\mathsf{H}}_{\boldsymbol{k}} \cdot \boldsymbol{\varPhi}_{\boldsymbol{k}}
$,
where $\boldsymbol{\varPhi}^\text{T}_{\boldsymbol{k}} = (a_{1,\boldsymbol{k}}, a_{2,\boldsymbol{k}})$ is built from the Fourier transformed Holstein-Primakoff bosons associated with the two sublattices of the honeycomb.
Relying on the well-established analysis of two-level systems \cite{Sticlet2012}, the Hamilton kernel
\begin{align}
	\tilde{\mathsf{H}}_{\boldsymbol{k}} = d^0_{\boldsymbol{k}} \sigma_0 + \boldsymbol{d}_{\boldsymbol{k}} \cdot \boldsymbol{\sigma}
\end{align}
is expanded in terms of Pauli matrices $\sigma_i$ ($i=0,1,2,3$), where $\sigma_0$ is the  $2\times 2$  unit matrix and $\boldsymbol{\sigma}^\text{T} = (\sigma_1,\sigma_2, \sigma_3)$. Its eigenvalues read
\begin{align}
	\varepsilon_{\boldsymbol{k},\pm} = d^0_{\boldsymbol{k}} \pm | \boldsymbol{d}_{\boldsymbol{k}} |.
\end{align}
As far as topology is concerned, $d^0_{\boldsymbol{k}} = 3JS+B$ is an irrelevant offset and the crucial information is encoded in the vector $\boldsymbol{d}_{\boldsymbol{k}}$ that determines the winding number
\begin{align}
	w = \frac{1}{4 \pi} \int_\text{BZ} \frac{\boldsymbol{d}_{\boldsymbol{k}}}{|\boldsymbol{d}_{\boldsymbol{k}}|^3} \cdot \left( \frac{\partial \boldsymbol{d}_{\boldsymbol{k}}}{\partial k_x} \times \frac{\partial \boldsymbol{d}_{\boldsymbol{k}}}{\partial k_y}\right)
	\mathrm{d}k_x \mathrm{d}k_y.
	\label{eq:windingnumberintegral}
\end{align}
The integration is over the entire Brillouin zone (BZ). The winding number measures how often $\boldsymbol{d}_{\boldsymbol{k}}$ wraps around the unit sphere. Using the explicit expression
\begin{align}
	\boldsymbol{d}_{\boldsymbol{k}}
	=
	\sum_{i=1}^3
	\begin{pmatrix}
		-J S \cos( \boldsymbol k \cdot \boldsymbol \delta_i ) \\
		 J S \sin( \boldsymbol k \cdot \boldsymbol \delta_i ) \\
		 2DS \sin( \boldsymbol k \cdot \boldsymbol \tau_i )
	\end{pmatrix},
\end{align}
where the vectors to nearest and second-nearest neighbors are given by
\begin{align}
	\boldsymbol{\delta}_1 &= (\sqrt{3}/2,1/2), \\
	\boldsymbol{\delta}_2 &= (-\sqrt{3}/2,1/2), \\
	\boldsymbol{\delta}_3 &= (0,-1),
\end{align}
and
\begin{align}
	\boldsymbol{\tau}_1 &= (\sqrt{3},0), \\
	\boldsymbol{\tau}_2 &= (-\sqrt{3}/2,3/2), \\
	\boldsymbol{\tau}_3 &= (-\sqrt{3}/2,-3/2),
\end{align}
respectively, the integral \eqref{eq:windingnumberintegral} may be evaluated numerically. Alternatively, one may reexpress the winding number in terms of the sign of the mass term $d^3_{\boldsymbol k}$ at the $K$ and $K'$ points of the Brillouin zone as\cite{Sticlet2012, Fruchart2013}
\begin{align}
	w = \frac{1}{2} \left[ \text{sgn}(d^3_{\boldsymbol K'}) - \text{sgn}(d^3_{\boldsymbol K}) \right].
\end{align}
Using $\boldsymbol K = (-4\pi/(3\sqrt{3}),0)$ and $\boldsymbol K' = - \boldsymbol K$, one obtains
\begin{align}
	d^3_{\boldsymbol K} = -d^3_{\boldsymbol K'} = 3 \sqrt{3} D S
\end{align}
and arrives at
\begin{align}
	w = - \text{sgn}( D ).
\end{align}
In the stack of honeycomb layers considered in the main text, the sign of DMI alternates between adjacent layers. Hence, the winding number of the $l$th layer reads $w^{(l)} = (-1)^l \text{sgn}( D )$.

For a ferromagnetic state pointing along the negative $z$ direction, one finds $w = \text{sgn}( D )$ because a reversal of the magnetization acts like a reversal of time that flips the chirality of the edge modes. Denoting the ferromagnetic order by $m = \pm 1$, with $m=+1$ ($m=-1$) referring to polarization along the positive (negative) $z$ direction, the winding number of each layer reads $w^{(l)} = (-1)^l \text{sgn}( m D )$.

\section{Second-order topology for inversion symmetric samples}
A special situation is found for samples that hold spatial symmetries, here, inversion symmetry, as present in Figs.~\ref{fig:HingeModes}(a,g). A domain wall necessarily is accompanied by another domain wall supporting a counter-propagating mode at the symmetry-related position. The two positions are separated by half of the sample's circumference, the largest possible spatial distance, maximally suppressing the hybridization of counter-propagating chiral hinge modes. (We assumed a convex sample cross-section.) In this spatially symmetric case, a pair of chiral hinge magnons is dictated by a bulk-hinge correspondence \cite{Takahashi2020}, a concept recently developed in the field of higher-order topological phases
\cite{Benalcazar2017, Benalcazar2017PRL, Langbehn2017, Song2017, Schindler2018, Geier2018, Kooi2018, Khalaf2018, Fang2019, Trifunovic2019}.

Mathematically, this is shown as follows (details are laid out in the SI \cite{Supplement}).
If the sample holds inversion symmetry, the Hamiltonian's Fourier kernel $\mathsf{H}_{\boldsymbol{k}}$ commutes with the parity operator $\mathsf{U}$ at time-reversal invariant momenta (TRIM) $\boldsymbol{\varGamma}_{abc} = (a \boldsymbol{g}_1 + b \boldsymbol{g}_2 + c \boldsymbol{g}_3)/2$, with $a,b,c \in \{0,1\}$ and the $\boldsymbol{g}_i$'s ($i=1,2,3$) being primitive reciprocal lattice vectors. Hence, at $\boldsymbol{k} = \boldsymbol{\varGamma}_{abc}$, $\mathsf{H}_{\boldsymbol{k}}$ and $\mathsf{U}$ share eigenvectors, with energy $\varepsilon_{n,\boldsymbol{\varGamma}_{abc}}$ and parity eigenvalues $p_{n,\boldsymbol{\varGamma}_{abc}} = \pm 1$\cite{Fu2007}. The latter enter the $\mathbb{Z}_4$ symmetry indicator \cite{Po2017, Bradlyn2017, Ono2018, Takahashi2020}
\begin{align}
    \mu_1
    =
    -\sum_{a = 0,1} \sum_{b = 0,1} \sum_{c = 0,1} n_-(\boldsymbol{\varGamma}_{abc}) \text{ mod } 4,
    \label{eq:mu1}
\end{align}
where $n_-(\boldsymbol{\varGamma}_{abc})$ is the number of \emph{negative} parity eigenvalues among the \emph{lowest two bands} at TRIM $\boldsymbol{\varGamma}_{abc}$.
An even (odd) $\mu_1$ indicates a band gap (Weyl points) between the second and third band; the gap is either trivial ($\mu_1 = 0$) or nontrivial ($\mu_1 = 2$)\cite{Hughes2011, Turner2012, Ono2018}.

With the derivation laid out in the SI \cite{Supplement}, we obtain the following second-order topological phase diagram. For the physically most relevant scenario of weakly coupled layers $j_\perp < \frac{1}{2}$, nontrivial second-order topology is found ($\mu_1 = 2$). By virtue of the bulk-hinge correspondence, hinge states are guaranteed if the finite sample holds inversion symmetry. Stronger interlayer coupling, $\frac{1}{2} < j_\perp < \frac{3}{2}$, causes a semimetallic phase with Weyl magnons\cite{Li2016, Mook2016, Su2017, Zyuzin2018} ($\mu_1 = 3$), which we study in more detail in the SI \cite{Supplement}. Even stronger coupling $\frac{3}{2} < j_\perp$ stabilizes a topologically trivial insulating phase ($\mu_1 = 0$) without hinge modes.

However, we stress that the existence of chiral hinge magnons does not require inversion or any other crystalline symmetry. In the SI we explicitly show that the hinge modes are robust against inversion-symmetry violating perturbations \cite{Supplement}.

\section{Atomistic spin dynamics simulations}
Due to the semiclassical nature of the harmonic magnon theory, the nontrivial topology of spin waves can be captured by classical spin dynamics. It is based on the equation of motion $\hbar \dot{\boldsymbol{S}}_i(t) = - \boldsymbol{S}_i(t) \times \boldsymbol{B}_i(t)$, describing the precession of each spin vector $\boldsymbol{S}_i$ in the effective magnetic field $\boldsymbol{B}_i$ due to its neighbors. For simplicity, Gilbert damping is neglected and temperature is set to zero.

Starting from the fully polarized ferromagnetic ground state (along the $z$ direction), we apply a dynamic magnetic field $\boldsymbol{b}_r = b_0 \sin( t \varepsilon_\text{ex} / \hbar )\hat{\boldsymbol{x}}$ to a single spin (index $r$); $\hat{\boldsymbol x}$ is a unit vector along the $x$ direction. It causes a coherent excitation of magnons at energy $\varepsilon_\text{ex}$, provided they have finite probability density at site $r$. A small amplitude $b_0 \ll 1$ is chosen to avoid nonlinear dynamics (which correspond to magnon-magnon interactions). To trace the excitation, we measure the discrete amplitude $A_i(t) = \sqrt{ [S_i^x(t)]^2 + [S_i^y(t)]^2 }$ of each spin upon numerical integration of the equation of motion. For plotting, we convert $A_i(t)$ into a continuous density $A(t,\boldsymbol{r})$, with $\boldsymbol{r}$ denoting the position in the finite sample.

We consider a finite-sized sample with $30$ or $31$ layers of $40 \times 40$ honeycomb unit cells with ``compensated'' boundaries (see SI for explanation \cite{Supplement}). The cross section of the pillar is a parallelogram with all-zigzag termination, hosting chiral hinge magnons at the two opposite obtuse corners [cf.~Figs.~\ref{fig:bz-and-bands}(e) and \ref{fig:HingeModes}(b,c)]. We set $\varepsilon_\text{ex}/(JS) = 3.4$, which is right in the middle of the global band gap [cf.~Fig.~\ref{fig:bz-and-bands}(d)].

\section{Implementation of disorder}
We add random magnetic fields $b_i^z$, drawn from a uniform distribution $[-\frac{\beta}{2},\frac{\beta}{2}]$, to all spins. Within linear spin-wave theory, the $b_i^z$'s enter the main diagonal of the Hamilton matrix, resembling chemical-potential disorder known from electronic disorder studies. 
The disorder strength is measured by its standard deviation $\sigma_b = \beta /(2 \sqrt{3})$.

\section{Effects of finite temperature}
The spin Hamiltonian \eqref{eq:hamiltonian} is $U(1)$ symmetric about the $z$ direction. Hence, number nonconserving interactions and spontaneous magnon decay are ruled out\cite{Zhitomirsky2013}, rendering number conserving four-magnon interactions the leading-order many-body perturbation. Their contribution to magnon damping is frozen out at zero temperature, but it renormalizes the magnon energy and damping at finite temperature \cite{Dyson1956}. 

At order $1/S$, the Hartree-like contribution (a Feynman diagram with a single four-magnon vertex) causes a purely real renormalization, which uniformly shifts the magnon energies downwards in energy. For a single honeycomb layer, this effect was already studied and found to scale with $T^2$, with the exponent determined by the dimension of the magnet\cite{Pershoguba2018}. Extending this analysis to three dimensions, we expect the scaling $T^{5/2}$. The uniform compression of the magnon spectrum leads to a reduction of the group velocities, implying that the chiral hinge magnons slow down as temperature increases.

Complex self-energies appear first at order $1/S^2$ and cause additional magnon damping, which also scales with $T^2$ in two dimensions\cite{Pershoguba2018} and, hence, with $T^{5/2}$ in three dimensions. The integrity of the chiral hinge modes is jeopardized when their temperature-induced lifetime broadening becomes as large as the surface band gap they cross. This condition defines a temperature $T'$, below which the device should be operated. The larger the surface gap, the larger $T'$. 

The above considerations apply to the case of zero magnetic field (and also zero easy-axis anisotropy). Both finite fields and easy-axis anisotropies shift the magnon spectrum uniformly towards higher energies, exponentially freezing out thermal effects. Hence, thermal effects can be systematically suppressed by external control.

\bibliographystyle{naturemag}
\bibliography{short,newrefs}

\begin{thebibliography}{10}
\expandafter\ifx\csname url\endcsname\relax
  \def\url#1{\texttt{#1}}\fi
\expandafter\ifx\csname urlprefix\endcsname\relax\def\urlprefix{URL }\fi
\providecommand{\bibinfo}[2]{#2}
\providecommand{\eprint}[2][]{\url{#2}}

\bibitem{Klitzing1986}
\bibinfo{author}{von Klitzing, K.}
\newblock The quantized {H}all effect.
\newblock \emph{\bibinfo{journal}{Rev. Mod. Phys.}}
  \textbf{\bibinfo{volume}{58}}, \bibinfo{pages}{519--531}
  (\bibinfo{year}{1986}).
\newblock \urlprefix\url{http://link.aps.org/doi/10.1103/RevModPhys.58.519}.

\bibitem{pankratov1987supersymmetric}
\bibinfo{author}{Pankratov, O.~A.}
\newblock Supersymmetric inhomogeneous semiconductor structures and the nature
  of a parity anomaly in (2+1) electrodynamics.
\newblock \emph{\bibinfo{journal}{Physics letters A}}
  \textbf{\bibinfo{volume}{121}}, \bibinfo{pages}{360--366}
  (\bibinfo{year}{1987}).

\bibitem{Haldane1988}
\bibinfo{author}{Haldane, F. D.~M.}
\newblock Model for a Quantum Hall Effect without Landau Levels:
  Condensed-Matter Realization of the ``Parity Anomaly''.
\newblock \emph{\bibinfo{journal}{Phys.\ Rev.\ Lett.}}
  \textbf{\bibinfo{volume}{61}}, \bibinfo{pages}{2015--2018}
  (\bibinfo{year}{1988}).
\newblock \urlprefix\url{http://dx.doi.org/10.1103/PhysRevLett.61.2015}.

\bibitem{Zhang2013}
\bibinfo{author}{Zhang, L.}, \bibinfo{author}{Ren, J.}, \bibinfo{author}{Wang,
  J.-S.} \& \bibinfo{author}{Li, B.}
\newblock Topological magnon insulator in insulating ferromagnet.
\newblock \emph{\bibinfo{journal}{Phys.\ Rev.\ B}}
  \textbf{\bibinfo{volume}{87}}, \bibinfo{pages}{144101}
  (\bibinfo{year}{2013}).
\newblock \urlprefix\url{http://dx.doi.org/10.1103/PhysRevB.87.144101}.

\bibitem{Shindou13}
\bibinfo{author}{Shindou, R.}, \bibinfo{author}{Matsumoto, R.},
  \bibinfo{author}{Murakami, S.} \& \bibinfo{author}{Ohe, J.-i.}
\newblock Topological chiral magnonic edge mode in a magnonic crystal.
\newblock \emph{\bibinfo{journal}{Phys. Rev. B}} \textbf{\bibinfo{volume}{87}},
  \bibinfo{pages}{174427} (\bibinfo{year}{2013}).
\newblock \urlprefix\url{http://link.aps.org/doi/10.1103/PhysRevB.87.174427}.

\bibitem{Mook14b}
\bibinfo{author}{Mook, A.}, \bibinfo{author}{Henk, J.} \&
  \bibinfo{author}{Mertig, I.}
\newblock Edge states in topological magnon insulators.
\newblock \emph{\bibinfo{journal}{Phys. Rev. B}} \textbf{\bibinfo{volume}{90}},
  \bibinfo{pages}{024412} (\bibinfo{year}{2014}).
\newblock \urlprefix\url{http://link.aps.org/doi/10.1103/PhysRevB.90.024412}.

\bibitem{Molina2016}
\bibinfo{author}{Rold\'{a}n-Molina, A.}, \bibinfo{author}{Nunez, A.~S.} \&
  \bibinfo{author}{Fern\'{a}ndez-Rossier, J.}
\newblock Topological spin waves in the atomic-scale magnetic skyrmion crystal.
\newblock \emph{\bibinfo{journal}{New Journal of Physics}}
  \textbf{\bibinfo{volume}{18}}, \bibinfo{pages}{045015}
  (\bibinfo{year}{2016}).
\newblock \urlprefix\url{http://stacks.iop.org/1367-2630/18/i=4/a=045015}.

\bibitem{Nakata2017QSHE}
\bibinfo{author}{Nakata, K.}, \bibinfo{author}{Kim, S.~K.},
  \bibinfo{author}{Klinovaja, J.} \& \bibinfo{author}{Loss, D.}
\newblock Magnonic topological insulators in antiferromagnets.
\newblock \emph{\bibinfo{journal}{Phys. Rev. B}} \textbf{\bibinfo{volume}{96}},
  \bibinfo{pages}{224414} (\bibinfo{year}{2017}).
\newblock \urlprefix\url{https://link.aps.org/doi/10.1103/PhysRevB.96.224414}.

\bibitem{Diaz2019}
\bibinfo{author}{D\'{\i}az, S.~A.}, \bibinfo{author}{Klinovaja, J.} \&
  \bibinfo{author}{Loss, D.}
\newblock Topological Magnons and Edge States in Antiferromagnetic Skyrmion
  Crystals.
\newblock \emph{\bibinfo{journal}{Phys. Rev. Lett.}}
  \textbf{\bibinfo{volume}{122}}, \bibinfo{pages}{187203}
  (\bibinfo{year}{2019}).
\newblock
  \urlprefix\url{https://link.aps.org/doi/10.1103/PhysRevLett.122.187203}.

\bibitem{Meier2003}
\bibinfo{author}{Meier, F.} \& \bibinfo{author}{Loss, D.}
\newblock Magnetization Transport and Quantized Spin Conductance.
\newblock \emph{\bibinfo{journal}{Phys. Rev. Lett.}}
  \textbf{\bibinfo{volume}{90}}, \bibinfo{pages}{167204}
  (\bibinfo{year}{2003}).
\newblock
  \urlprefix\url{http://link.aps.org/doi/10.1103/PhysRevLett.90.167204}.

\bibitem{Katsura2010}
\bibinfo{author}{Katsura, H.}, \bibinfo{author}{Nagaosa, N.} \&
  \bibinfo{author}{Lee, P.~A.}
\newblock Theory of the Thermal Hall Effect in Quantum Magnets.
\newblock \emph{\bibinfo{journal}{Phys.\ Rev.\ Lett.}}
  \textbf{\bibinfo{volume}{104}}, \bibinfo{pages}{066403}
  (\bibinfo{year}{2010}).
\newblock \urlprefix\url{http://dx.doi.org/10.1103/PhysRevLett.104.066403}.

\bibitem{Matsumoto2011a}
\bibinfo{author}{Matsumoto, R.} \& \bibinfo{author}{Murakami, S.}
\newblock Theoretical Prediction of a Rotating Magnon Wave Packet in
  Ferromagnets.
\newblock \emph{\bibinfo{journal}{Phys.\ Rev.\ Lett.}}
  \textbf{\bibinfo{volume}{106}}, \bibinfo{pages}{197202}
  (\bibinfo{year}{2011}).
\newblock \urlprefix\url{http://dx.doi.org/10.1103/PhysRevLett.106.197202}.

\bibitem{Hoogdalem2013}
\bibinfo{author}{van Hoogdalem, K.~A.}, \bibinfo{author}{Tserkovnyak, Y.} \&
  \bibinfo{author}{Loss, D.}
\newblock Magnetic texture-induced thermal Hall effects.
\newblock \emph{\bibinfo{journal}{Phys. Rev. B}} \textbf{\bibinfo{volume}{87}},
  \bibinfo{pages}{024402} (\bibinfo{year}{2013}).
\newblock \urlprefix\url{http://link.aps.org/doi/10.1103/PhysRevB.87.024402}.

\bibitem{Mook14a}
\bibinfo{author}{Mook, A.}, \bibinfo{author}{Henk, J.} \&
  \bibinfo{author}{Mertig, I.}
\newblock Magnon Hall effect and topology in kagome lattices: A theoretical
  investigation.
\newblock \emph{\bibinfo{journal}{Phys. Rev. B}} \textbf{\bibinfo{volume}{89}},
  \bibinfo{pages}{134409} (\bibinfo{year}{2014}).
\newblock \urlprefix\url{http://link.aps.org/doi/10.1103/PhysRevB.89.134409}.

\bibitem{Nakata2016}
\bibinfo{author}{Nakata, K.}, \bibinfo{author}{Klinovaja, J.} \&
  \bibinfo{author}{Loss, D.}
\newblock Magnonic quantum Hall effect and Wiedemann-Franz law.
\newblock \emph{\bibinfo{journal}{Phys. Rev. B}} \textbf{\bibinfo{volume}{95}},
  \bibinfo{pages}{125429} (\bibinfo{year}{2017}).
\newblock \urlprefix\url{https://link.aps.org/doi/10.1103/PhysRevB.95.125429}.

\bibitem{Mook2018}
\bibinfo{author}{Mook, A.}, \bibinfo{author}{G\"obel, B.},
  \bibinfo{author}{Henk, J.} \& \bibinfo{author}{Mertig, I.}
\newblock Taking an electron-magnon duality shortcut from electron to magnon
  transport.
\newblock \emph{\bibinfo{journal}{Phys. Rev. B}} \textbf{\bibinfo{volume}{97}},
  \bibinfo{pages}{140401} (\bibinfo{year}{2018}).
\newblock \urlprefix\url{https://link.aps.org/doi/10.1103/PhysRevB.97.140401}.

\bibitem{Kim2019}
\bibinfo{author}{Kim, S.~K.}, \bibinfo{author}{Nakata, K.},
  \bibinfo{author}{Loss, D.} \& \bibinfo{author}{Tserkovnyak, Y.}
\newblock Tunable Magnonic Thermal Hall Effect in Skyrmion Crystal Phases of
  Ferrimagnets.
\newblock \emph{\bibinfo{journal}{Phys. Rev. Lett.}}
  \textbf{\bibinfo{volume}{122}}, \bibinfo{pages}{057204}
  (\bibinfo{year}{2019}).
\newblock
  \urlprefix\url{https://link.aps.org/doi/10.1103/PhysRevLett.122.057204}.

\bibitem{Chumak2015}
\bibinfo{author}{Chumak, A.~V.}, \bibinfo{author}{Vasyuchka, V.~I.},
  \bibinfo{author}{Serga, A.~A.} \& \bibinfo{author}{Hillebrands, B.}
\newblock Magnon spintronics.
\newblock \emph{\bibinfo{journal}{Nature Physics}}
  \textbf{\bibinfo{volume}{11}}, \bibinfo{pages}{453--461}
  (\bibinfo{year}{2015}).
\newblock \urlprefix\url{https://doi.org/10.1038/nphys3347}.

\bibitem{Wang2018TopMagnonics}
\bibinfo{author}{Wang, X.~S.}, \bibinfo{author}{Zhang, H.~W.} \&
  \bibinfo{author}{Wang, X.~R.}
\newblock Topological Magnonics: A Paradigm for Spin-Wave Manipulation and
  Device Design.
\newblock \emph{\bibinfo{journal}{Phys. Rev. Applied}}
  \textbf{\bibinfo{volume}{9}}, \bibinfo{pages}{024029} (\bibinfo{year}{2018}).
\newblock
  \urlprefix\url{https://link.aps.org/doi/10.1103/PhysRevApplied.9.024029}.

\bibitem{Yamamoto2019}
\bibinfo{author}{Yamamoto, K.} \emph{et~al.}
\newblock Topological Characterization of Classical Waves: The Topological
  Origin of Magnetostatic Surface Spin Waves.
\newblock \emph{\bibinfo{journal}{Phys. Rev. Lett.}}
  \textbf{\bibinfo{volume}{122}}, \bibinfo{pages}{217201}
  (\bibinfo{year}{2019}).
\newblock
  \urlprefix\url{https://link.aps.org/doi/10.1103/PhysRevLett.122.217201}.

\bibitem{Diaz2020}
\bibinfo{author}{D\'{\i}az, S.~A.}, \bibinfo{author}{Hirosawa, T.},
  \bibinfo{author}{Klinovaja, J.} \& \bibinfo{author}{Loss, D.}
\newblock Chiral magnonic edge states in ferromagnetic skyrmion crystals
  controlled by magnetic fields.
\newblock \emph{\bibinfo{journal}{Phys. Rev. Research}}
  \textbf{\bibinfo{volume}{2}}, \bibinfo{pages}{013231} (\bibinfo{year}{2020}).
\newblock
  \urlprefix\url{https://link.aps.org/doi/10.1103/PhysRevResearch.2.013231}.

\bibitem{Aguilera2020}
\bibinfo{author}{Aguilera, E.}, \bibinfo{author}{Jaeschke-Ubiergo, R.},
  \bibinfo{author}{Vidal-Silva, N.}, \bibinfo{author}{Torres, L. E. F.~F.} \&
  \bibinfo{author}{Nunez, A.~S.}
\newblock Topological magnonics in the two-dimensional van der Waals magnet
  ${\mathrm{CrI}}_{3}$.
\newblock \emph{\bibinfo{journal}{Phys. Rev. B}}
  \textbf{\bibinfo{volume}{102}}, \bibinfo{pages}{024409}
  (\bibinfo{year}{2020}).
\newblock \urlprefix\url{https://link.aps.org/doi/10.1103/PhysRevB.102.024409}.

\bibitem{Topol2006}
\bibinfo{author}{{Topol}, A.~W.} \emph{et~al.}
\newblock Three-dimensional integrated circuits.
\newblock \emph{\bibinfo{journal}{IBM Journal of Research and Development}}
  \textbf{\bibinfo{volume}{50}}, \bibinfo{pages}{491--506}
  (\bibinfo{year}{2006}).

\bibitem{Benalcazar2017}
\bibinfo{author}{Benalcazar, W.~A.}, \bibinfo{author}{Bernevig, B.~A.} \&
  \bibinfo{author}{Hughes, T.~L.}
\newblock Quantized electric multipole insulators.
\newblock \emph{\bibinfo{journal}{Science}} \textbf{\bibinfo{volume}{357}},
  \bibinfo{pages}{61--66} (\bibinfo{year}{2017}).
\newblock \urlprefix\url{https://doi.org/10.1126/science.aah6442}.

\bibitem{Schindler2018}
\bibinfo{author}{Schindler, F.} \emph{et~al.}
\newblock Higher-order topology in bismuth.
\newblock \emph{\bibinfo{journal}{Nature Physics}}
  \textbf{\bibinfo{volume}{14}}, \bibinfo{pages}{918--924}
  (\bibinfo{year}{2018}).
\newblock \urlprefix\url{https://doi.org/10.1038/s41567-018-0224-7}.

\bibitem{Li2019HOTISoliton}
\bibinfo{author}{Li, Z.}, \bibinfo{author}{Cao, Y.}, \bibinfo{author}{Yan, P.}
  \& \bibinfo{author}{Wang, X.}
\newblock Higher-order topological solitonic insulators.
\newblock \emph{\bibinfo{journal}{npj Computational Materials}}
  \textbf{\bibinfo{volume}{5}}, \bibinfo{pages}{107} (\bibinfo{year}{2019}).
\newblock \urlprefix\url{https://doi.org/10.1038/s41524-019-0246-4}.

\bibitem{Sil2020}
\bibinfo{author}{Sil, A.} \& \bibinfo{author}{Ghosh, A.~K.}
\newblock First and second order topological phases on ferromagnetic breathing
  kagome lattice.
\newblock \emph{\bibinfo{journal}{Journal of Physics: Condensed Matter}}
  \textbf{\bibinfo{volume}{32}}, \bibinfo{pages}{205601}
  (\bibinfo{year}{2020}).
\newblock \urlprefix\url{https://doi.org/10.1088%2F1361-648x%2Fab6f8b}.

\bibitem{Hirosawa2020TopQuadrupole}
\bibinfo{author}{Hirosawa, T.}, \bibinfo{author}{D\'{\i}az, S.~A.},
  \bibinfo{author}{Klinovaja, J.} \& \bibinfo{author}{Loss, D.}
\newblock Magnonic Quadrupole Topological Insulator in Antiskyrmion Crystals.
\newblock \emph{\bibinfo{journal}{arXiv:2005.05884}}  (\bibinfo{year}{2020}).

\bibitem{Owerre2016a}
\bibinfo{author}{Owerre, S.~A.}
\newblock A first theoretical realization of honeycomb topological magnon
  insulator.
\newblock \emph{\bibinfo{journal}{J. Phys.: Condens. Matter}}
  \textbf{\bibinfo{volume}{28}}, \bibinfo{pages}{386001}
  (\bibinfo{year}{2016}).
\newblock \urlprefix\url{http://dx.doi.org/10.1088/0953-8984/28/38/386001}.

\bibitem{Dzyaloshinsky58}
\bibinfo{author}{Dzyaloshinsky, I.}
\newblock A thermodynamic theory of ``weak'' ferromagnetism of
  antiferromagnetics.
\newblock \emph{\bibinfo{journal}{J. Phys.\ Chem.\ Sol.}}
  \textbf{\bibinfo{volume}{4}}, \bibinfo{pages}{241} (\bibinfo{year}{1958}).

\bibitem{Moriya60}
\bibinfo{author}{Moriya, T.}
\newblock Anisotropic Superexchange Interaction and Weak Ferromagnetism.
\newblock \emph{\bibinfo{journal}{Phys.\ Rev.}} \textbf{\bibinfo{volume}{120}},
  \bibinfo{pages}{91} (\bibinfo{year}{1960}).

\bibitem{Zhang2019vanderWaalsReview}
\bibinfo{author}{Zhang, W.}, \bibinfo{author}{Wong, P. K.~J.},
  \bibinfo{author}{Zhu, R.} \& \bibinfo{author}{Wee, A. T.~S.}
\newblock Van der Waals magnets: Wonder building blocks for two-dimensional
  spintronics?
\newblock \emph{\bibinfo{journal}{{InfoMat}}} \textbf{\bibinfo{volume}{1}},
  \bibinfo{pages}{479--495} (\bibinfo{year}{2019}).
\newblock \urlprefix\url{https://doi.org/10.1002/inf2.12048}.

\bibitem{Teo2008}
\bibinfo{author}{Teo, J. C.~Y.}, \bibinfo{author}{Fu, L.} \&
  \bibinfo{author}{Kane, C.~L.}
\newblock Surface states and topological invariants in three-dimensional
  topological insulators: Application to
  ${\text{Bi}}_{1\ensuremath{-}x}{\text{Sb}}_{x}$.
\newblock \emph{\bibinfo{journal}{Phys. Rev. B}} \textbf{\bibinfo{volume}{78}},
  \bibinfo{pages}{045426} (\bibinfo{year}{2008}).
\newblock \urlprefix\url{https://link.aps.org/doi/10.1103/PhysRevB.78.045426}.

\bibitem{Fulga2016}
\bibinfo{author}{Fulga, I.~C.}, \bibinfo{author}{Avraham, N.},
  \bibinfo{author}{Beidenkopf, H.} \& \bibinfo{author}{Stern, A.}
\newblock Coupled-layer description of topological crystalline insulators.
\newblock \emph{\bibinfo{journal}{Phys. Rev. B}} \textbf{\bibinfo{volume}{94}},
  \bibinfo{pages}{125405} (\bibinfo{year}{2016}).
\newblock \urlprefix\url{https://link.aps.org/doi/10.1103/PhysRevB.94.125405}.

\bibitem{Supplement}
\bibinfo{note}{The Supplementary Information contains accompanying information
  on the following points: (1) linear spin-wave theory of the model
  Hamiltonian, (2) other non-second-order topological phases of the model
  Hamiltonian, (3) the influence of the sign of $\delta j_\perp$, (4)
  calculation of second-order symmetry indicators, (5) spin dynamics
  simulations, (6) the effect of ``uncompensated'' boundaries, and (7) the
  effects of inversion-symmetry breaking perturbations.. The Supplementary
  Information also includes online movies of spin dynamics simulations and
  additional references. URL: [inserted by publisher]}.

\bibitem{CastroNeto2009}
\bibinfo{author}{Castro~Neto, A.~H.}, \bibinfo{author}{Guinea, F.},
  \bibinfo{author}{Peres, N. M.~R.}, \bibinfo{author}{Novoselov, K.~S.} \&
  \bibinfo{author}{Geim, A.~K.}
\newblock The electronic properties of graphene.
\newblock \emph{\bibinfo{journal}{Rev. Mod. Phys.}}
  \textbf{\bibinfo{volume}{81}}, \bibinfo{pages}{109--162}
  (\bibinfo{year}{2009}).
\newblock \urlprefix\url{https://link.aps.org/doi/10.1103/RevModPhys.81.109}.

\bibitem{Su1979}
\bibinfo{author}{Su, W.~P.}, \bibinfo{author}{Schrieffer, J.~R.} \&
  \bibinfo{author}{Heeger, A.~J.}
\newblock Solitons in Polyacetylene.
\newblock \emph{\bibinfo{journal}{Phys. Rev. Lett.}}
  \textbf{\bibinfo{volume}{42}}, \bibinfo{pages}{1698--1701}
  (\bibinfo{year}{1979}).
\newblock \urlprefix\url{https://link.aps.org/doi/10.1103/PhysRevLett.42.1698}.

\bibitem{Takahashi2020}
\bibinfo{author}{Takahashi, R.}, \bibinfo{author}{Tanaka, Y.} \&
  \bibinfo{author}{Murakami, S.}
\newblock Bulk-edge and bulk-hinge correspondence in inversion-symmetric
  insulators.
\newblock \emph{\bibinfo{journal}{Phys. Rev. Research}}
  \textbf{\bibinfo{volume}{2}}, \bibinfo{pages}{013300} (\bibinfo{year}{2020}).
\newblock
  \urlprefix\url{https://link.aps.org/doi/10.1103/PhysRevResearch.2.013300}.

\bibitem{Brataas2008}
\bibinfo{author}{Brataas, A.}, \bibinfo{author}{Tserkovnyak, Y.} \&
  \bibinfo{author}{Bauer, G. E.~W.}
\newblock Scattering Theory of Gilbert Damping.
\newblock \emph{\bibinfo{journal}{Phys. Rev. Lett.}}
  \textbf{\bibinfo{volume}{101}}, \bibinfo{pages}{037207}
  (\bibinfo{year}{2008}).
\newblock
  \urlprefix\url{https://link.aps.org/doi/10.1103/PhysRevLett.101.037207}.

\bibitem{Chen2020}
\bibinfo{author}{Chen, L.} \emph{et~al.}
\newblock Magnetic anisotropy in ferromagnetic ${\mathrm{CrI}}_{3}$.
\newblock \emph{\bibinfo{journal}{Phys. Rev. B}}
  \textbf{\bibinfo{volume}{101}}, \bibinfo{pages}{134418}
  (\bibinfo{year}{2020}).
\newblock \urlprefix\url{https://link.aps.org/doi/10.1103/PhysRevB.101.134418}.

\bibitem{Qin2019Zakeri}
\bibinfo{author}{Qin, H.~J.}, \bibinfo{author}{Tsurkan, S.},
  \bibinfo{author}{Ernst, A.} \& \bibinfo{author}{Zakeri, K.}
\newblock Experimental Realization of Atomic-Scale Magnonic Crystals.
\newblock \emph{\bibinfo{journal}{Phys. Rev. Lett.}}
  \textbf{\bibinfo{volume}{123}}, \bibinfo{pages}{257202}
  (\bibinfo{year}{2019}).
\newblock
  \urlprefix\url{https://link.aps.org/doi/10.1103/PhysRevLett.123.257202}.

\bibitem{Tokmachev2018}
\bibinfo{author}{Tokmachev, A.~M.} \emph{et~al.}
\newblock Emerging two-dimensional ferromagnetism in silicene materials.
\newblock \emph{\bibinfo{journal}{Nature Communications}}
  \textbf{\bibinfo{volume}{9}} (\bibinfo{year}{2018}).
\newblock \urlprefix\url{https://doi.org/10.1038/s41467-018-04012-2}.

\bibitem{Huang2017}
\bibinfo{author}{Huang, B.} \emph{et~al.}
\newblock Layer-dependent ferromagnetism in a van der Waals crystal down to the
  monolayer limit.
\newblock \emph{\bibinfo{journal}{Nature}} \textbf{\bibinfo{volume}{546}},
  \bibinfo{pages}{270--273} (\bibinfo{year}{2017}).
\newblock \urlprefix\url{https://doi.org/10.1038/nature22391}.

\bibitem{Liu2018orgmag}
\bibinfo{author}{Liu, H.} \emph{et~al.}
\newblock Organic-based magnon spintronics.
\newblock \emph{\bibinfo{journal}{Nature Materials}}
  \textbf{\bibinfo{volume}{17}}, \bibinfo{pages}{308--312}
  (\bibinfo{year}{2018}).
\newblock \urlprefix\url{https://doi.org/10.1038/s41563-018-0035-3}.

\bibitem{Talatchian2020}
\bibinfo{author}{Talatchian, P.} \emph{et~al.}
\newblock Designing Large Arrays of Interacting Spin-Torque Nano-Oscillators
  for Microwave Information Processing.
\newblock \emph{\bibinfo{journal}{Phys. Rev. Applied}}
  \textbf{\bibinfo{volume}{13}}, \bibinfo{pages}{024073}
  (\bibinfo{year}{2020}).
\newblock
  \urlprefix\url{https://link.aps.org/doi/10.1103/PhysRevApplied.13.024073}.

\bibitem{Pulecio2014}
\bibinfo{author}{Pulecio, J.~F.}, \bibinfo{author}{Warnicke, P.},
  \bibinfo{author}{Pollard, S.~D.}, \bibinfo{author}{Arena, D.~A.} \&
  \bibinfo{author}{Zhu, Y.}
\newblock Coherence and modality of driven interlayer-coupled magnetic
  vortices.
\newblock \emph{\bibinfo{journal}{Nature Communications}}
  \textbf{\bibinfo{volume}{5}} (\bibinfo{year}{2014}).
\newblock \urlprefix\url{https://doi.org/10.1038/ncomms4760}.

\bibitem{Rusconi2019}
\bibinfo{author}{Rusconi, C.~C.}, \bibinfo{author}{Schuetz, M. J.~A.},
  \bibinfo{author}{Gieseler, J.}, \bibinfo{author}{Lukin, M.~D.} \&
  \bibinfo{author}{Romero-Isart, O.}
\newblock Hybrid architecture for engineering magnonic quantum networks.
\newblock \emph{\bibinfo{journal}{Phys. Rev. A}}
  \textbf{\bibinfo{volume}{100}}, \bibinfo{pages}{022343}
  (\bibinfo{year}{2019}).
\newblock \urlprefix\url{https://link.aps.org/doi/10.1103/PhysRevA.100.022343}.

\bibitem{Cai2019}
\bibinfo{author}{Cai, W.} \emph{et~al.}
\newblock Observation of Topological Magnon Insulator States in a
  Superconducting Circuit.
\newblock \emph{\bibinfo{journal}{Phys. Rev. Lett.}}
  \textbf{\bibinfo{volume}{123}}, \bibinfo{pages}{080501}
  (\bibinfo{year}{2019}).
\newblock
  \urlprefix\url{https://link.aps.org/doi/10.1103/PhysRevLett.123.080501}.

\bibitem{Li2018magcrys}
\bibinfo{author}{Li, Y.-M.}, \bibinfo{author}{Xiao, J.} \&
  \bibinfo{author}{Chang, K.}
\newblock Topological Magnon Modes in Patterned Ferrimagnetic Insulator Thin
  Films.
\newblock \emph{\bibinfo{journal}{Nano Letters}} \textbf{\bibinfo{volume}{18}},
  \bibinfo{pages}{3032--3037} (\bibinfo{year}{2018}).
\newblock \urlprefix\url{https://doi.org/10.1021/acs.nanolett.8b00492}.

\bibitem{Vogt2012}
\bibinfo{author}{Vogt, K.} \emph{et~al.}
\newblock Spin waves turning a corner.
\newblock \emph{\bibinfo{journal}{Applied Physics Letters}}
  \textbf{\bibinfo{volume}{101}}, \bibinfo{pages}{042410}
  (\bibinfo{year}{2012}).
\newblock \urlprefix\url{https://doi.org/10.1063/1.4738887}.

\bibitem{Gubbiotti2019}
\bibinfo{author}{Gubbiotti, G.}
\newblock \emph{\bibinfo{title}{Three-Dimensional Magnonics}}
  (\bibinfo{publisher}{Jenny Stanford Publishing}, \bibinfo{year}{2019}).

\bibitem{Holstein1940}
\bibinfo{author}{Holstein, T.} \& \bibinfo{author}{Primakoff, H.}
\newblock Field Dependence of the Intrinsic Domain Magnetization of a
  Ferromagnet.
\newblock \emph{\bibinfo{journal}{Phys. Rev.}} \textbf{\bibinfo{volume}{58}},
  \bibinfo{pages}{1098--1113} (\bibinfo{year}{1940}).
\newblock \urlprefix\url{http://link.aps.org/doi/10.1103/PhysRev.58.1098}.

\bibitem{Sticlet2012}
\bibinfo{author}{Sticlet, D.}, \bibinfo{author}{Pi\'echon, F.},
  \bibinfo{author}{Fuchs, J.-N.}, \bibinfo{author}{Kalugin, P.} \&
  \bibinfo{author}{Simon, P.}
\newblock Geometrical engineering of a two-band Chern insulator in two
  dimensions with arbitrary topological index.
\newblock \emph{\bibinfo{journal}{Phys. Rev. B}} \textbf{\bibinfo{volume}{85}},
  \bibinfo{pages}{165456} (\bibinfo{year}{2012}).
\newblock \urlprefix\url{https://link.aps.org/doi/10.1103/PhysRevB.85.165456}.

\bibitem{Fruchart2013}
\bibinfo{author}{Fruchart, M.} \& \bibinfo{author}{Carpentier, D.}
\newblock An introduction to topological insulators.
\newblock \emph{\bibinfo{journal}{Comptes Rendus Physique}}
  \textbf{\bibinfo{volume}{14}}, \bibinfo{pages}{779--815}
  (\bibinfo{year}{2013}).
\newblock \urlprefix\url{https://doi.org/10.1016/j.crhy.2013.09.013}.

\bibitem{Benalcazar2017PRL}
\bibinfo{author}{Benalcazar, W.~A.}, \bibinfo{author}{Bernevig, B.~A.} \&
  \bibinfo{author}{Hughes, T.~L.}
\newblock Electric multipole moments, topological multipole moment pumping, and
  chiral hinge states in crystalline insulators.
\newblock \emph{\bibinfo{journal}{Phys. Rev. B}} \textbf{\bibinfo{volume}{96}},
  \bibinfo{pages}{245115} (\bibinfo{year}{2017}).
\newblock \urlprefix\url{https://link.aps.org/doi/10.1103/PhysRevB.96.245115}.

\bibitem{Langbehn2017}
\bibinfo{author}{Langbehn, J.}, \bibinfo{author}{Peng, Y.},
  \bibinfo{author}{Trifunovic, L.}, \bibinfo{author}{von Oppen, F.} \&
  \bibinfo{author}{Brouwer, P.~W.}
\newblock Reflection-Symmetric Second-Order Topological Insulators and
  Superconductors.
\newblock \emph{\bibinfo{journal}{Phys. Rev. Lett.}}
  \textbf{\bibinfo{volume}{119}}, \bibinfo{pages}{246401}
  (\bibinfo{year}{2017}).
\newblock
  \urlprefix\url{https://link.aps.org/doi/10.1103/PhysRevLett.119.246401}.

\bibitem{Song2017}
\bibinfo{author}{Song, Z.}, \bibinfo{author}{Fang, Z.} \&
  \bibinfo{author}{Fang, C.}
\newblock $(d\ensuremath{-}2)$-Dimensional Edge States of Rotation Symmetry
  Protected Topological States.
\newblock \emph{\bibinfo{journal}{Phys. Rev. Lett.}}
  \textbf{\bibinfo{volume}{119}}, \bibinfo{pages}{246402}
  (\bibinfo{year}{2017}).
\newblock
  \urlprefix\url{https://link.aps.org/doi/10.1103/PhysRevLett.119.246402}.

\bibitem{Geier2018}
\bibinfo{author}{Geier, M.}, \bibinfo{author}{Trifunovic, L.},
  \bibinfo{author}{Hoskam, M.} \& \bibinfo{author}{Brouwer, P.~W.}
\newblock Second-order topological insulators and superconductors with an
  order-two crystalline symmetry.
\newblock \emph{\bibinfo{journal}{Phys. Rev. B}} \textbf{\bibinfo{volume}{97}},
  \bibinfo{pages}{205135} (\bibinfo{year}{2018}).
\newblock \urlprefix\url{https://link.aps.org/doi/10.1103/PhysRevB.97.205135}.

\bibitem{Kooi2018}
\bibinfo{author}{Kooi, S.~H.}, \bibinfo{author}{van Miert, G.} \&
  \bibinfo{author}{Ortix, C.}
\newblock Inversion-symmetry protected chiral hinge states in stacks of doped
  quantum Hall layers.
\newblock \emph{\bibinfo{journal}{Phys. Rev. B}} \textbf{\bibinfo{volume}{98}},
  \bibinfo{pages}{245102} (\bibinfo{year}{2018}).
\newblock \urlprefix\url{https://link.aps.org/doi/10.1103/PhysRevB.98.245102}.

\bibitem{Khalaf2018}
\bibinfo{author}{Khalaf, E.}, \bibinfo{author}{Po, H.~C.},
  \bibinfo{author}{Vishwanath, A.} \& \bibinfo{author}{Watanabe, H.}
\newblock Symmetry Indicators and Anomalous Surface States of Topological
  Crystalline Insulators.
\newblock \emph{\bibinfo{journal}{Phys. Rev. X}} \textbf{\bibinfo{volume}{8}},
  \bibinfo{pages}{031070} (\bibinfo{year}{2018}).
\newblock \urlprefix\url{https://link.aps.org/doi/10.1103/PhysRevX.8.031070}.

\bibitem{Fang2019}
\bibinfo{author}{Fang, C.} \& \bibinfo{author}{Fu, L.}
\newblock New classes of topological crystalline insulators having surface
  rotation anomaly.
\newblock \emph{\bibinfo{journal}{Science Advances}}
  \textbf{\bibinfo{volume}{5}}, \bibinfo{pages}{eaat2374}
  (\bibinfo{year}{2019}).
\newblock \urlprefix\url{https://doi.org/10.1126/sciadv.aat2374}.

\bibitem{Trifunovic2019}
\bibinfo{author}{Trifunovic, L.} \& \bibinfo{author}{Brouwer, P.~W.}
\newblock Higher-Order Bulk-Boundary Correspondence for Topological Crystalline
  Phases.
\newblock \emph{\bibinfo{journal}{Phys. Rev. X}} \textbf{\bibinfo{volume}{9}},
  \bibinfo{pages}{011012} (\bibinfo{year}{2019}).
\newblock \urlprefix\url{https://link.aps.org/doi/10.1103/PhysRevX.9.011012}.

\bibitem{Fu2007}
\bibinfo{author}{Fu, L.} \& \bibinfo{author}{Kane, C.~L.}
\newblock Topological insulators with inversion symmetry.
\newblock \emph{\bibinfo{journal}{Phys. Rev. B}} \textbf{\bibinfo{volume}{76}},
  \bibinfo{pages}{045302} (\bibinfo{year}{2007}).
\newblock \urlprefix\url{https://link.aps.org/doi/10.1103/PhysRevB.76.045302}.

\bibitem{Po2017}
\bibinfo{author}{Po, H.~C.}, \bibinfo{author}{Vishwanath, A.} \&
  \bibinfo{author}{Watanabe, H.}
\newblock Symmetry-based indicators of band topology in the 230 space groups.
\newblock \emph{\bibinfo{journal}{Nature Communications}}
  \textbf{\bibinfo{volume}{8}}, \bibinfo{pages}{50} (\bibinfo{year}{2017}).
\newblock \urlprefix\url{https://doi.org/10.1038/s41467-017-00133-2}.

\bibitem{Bradlyn2017}
\bibinfo{author}{Bradlyn, B.} \emph{et~al.}
\newblock Topological quantum chemistry.
\newblock \emph{\bibinfo{journal}{Nature}} \textbf{\bibinfo{volume}{547}},
  \bibinfo{pages}{298--305} (\bibinfo{year}{2017}).
\newblock \urlprefix\url{https://doi.org/10.1038/nature23268}.

\bibitem{Ono2018}
\bibinfo{author}{Ono, S.} \& \bibinfo{author}{Watanabe, H.}
\newblock Unified understanding of symmetry indicators for all internal
  symmetry classes.
\newblock \emph{\bibinfo{journal}{Phys. Rev. B}} \textbf{\bibinfo{volume}{98}},
  \bibinfo{pages}{115150} (\bibinfo{year}{2018}).
\newblock \urlprefix\url{https://link.aps.org/doi/10.1103/PhysRevB.98.115150}.

\bibitem{Hughes2011}
\bibinfo{author}{Hughes, T.~L.}, \bibinfo{author}{Prodan, E.} \&
  \bibinfo{author}{Bernevig, B.~A.}
\newblock Inversion-symmetric topological insulators.
\newblock \emph{\bibinfo{journal}{Phys. Rev. B}} \textbf{\bibinfo{volume}{83}},
  \bibinfo{pages}{245132} (\bibinfo{year}{2011}).
\newblock \urlprefix\url{https://link.aps.org/doi/10.1103/PhysRevB.83.245132}.

\bibitem{Turner2012}
\bibinfo{author}{Turner, A.~M.}, \bibinfo{author}{Zhang, Y.},
  \bibinfo{author}{Mong, R. S.~K.} \& \bibinfo{author}{Vishwanath, A.}
\newblock Quantized response and topology of magnetic insulators with inversion
  symmetry.
\newblock \emph{\bibinfo{journal}{Phys. Rev. B}} \textbf{\bibinfo{volume}{85}},
  \bibinfo{pages}{165120} (\bibinfo{year}{2012}).
\newblock \urlprefix\url{https://link.aps.org/doi/10.1103/PhysRevB.85.165120}.

\bibitem{Li2016}
\bibinfo{author}{Li, F.-Y.} \emph{et~al.}
\newblock Weyl magnons in breathing pyrochlore antiferromagnets.
\newblock \emph{\bibinfo{journal}{Nature Communications}}
  \textbf{\bibinfo{volume}{7}}, \bibinfo{pages}{12691} (\bibinfo{year}{2016}).
\newblock \urlprefix\url{http://dx.doi.org/10.1038/ncomms12691}.

\bibitem{Mook2016}
\bibinfo{author}{Mook, A.}, \bibinfo{author}{Henk, J.} \&
  \bibinfo{author}{Mertig, I.}
\newblock Tunable Magnon {W}eyl Points in Ferromagnetic Pyrochlores.
\newblock \emph{\bibinfo{journal}{Phys. Rev. Lett.}}
  \textbf{\bibinfo{volume}{117}}, \bibinfo{pages}{157204}
  (\bibinfo{year}{2016}).
\newblock
  \urlprefix\url{https://link.aps.org/doi/10.1103/PhysRevLett.117.157204}.

\bibitem{Su2017}
\bibinfo{author}{Su, Y.} \& \bibinfo{author}{Wang, X.~R.}
\newblock Chiral anomaly of Weyl magnons in stacked honeycomb ferromagnets.
\newblock \emph{\bibinfo{journal}{Phys. Rev. B}} \textbf{\bibinfo{volume}{96}},
  \bibinfo{pages}{104437} (\bibinfo{year}{2017}).
\newblock \urlprefix\url{https://link.aps.org/doi/10.1103/PhysRevB.96.104437}.

\bibitem{Zyuzin2018}
\bibinfo{author}{Zyuzin, V.~A.} \& \bibinfo{author}{Kovalev, A.~A.}
\newblock Spin Hall and Nernst effects of Weyl magnons.
\newblock \emph{\bibinfo{journal}{Phys. Rev. B}} \textbf{\bibinfo{volume}{97}},
  \bibinfo{pages}{174407} (\bibinfo{year}{2018}).
\newblock \urlprefix\url{https://link.aps.org/doi/10.1103/PhysRevB.97.174407}.

\bibitem{Zhitomirsky2013}
\bibinfo{author}{Zhitomirsky, M.~E.} \& \bibinfo{author}{Chernyshev, A.~L.}
\newblock Colloquium: Spontaneous magnon decays.
\newblock \emph{\bibinfo{journal}{Rev. Mod. Phys.}}
  \textbf{\bibinfo{volume}{85}}, \bibinfo{pages}{219--242}
  (\bibinfo{year}{2013}).
\newblock \urlprefix\url{https://link.aps.org/doi/10.1103/RevModPhys.85.219}.

\bibitem{Dyson1956}
\bibinfo{author}{Dyson, F.~J.}
\newblock General Theory of Spin-Wave Interactions.
\newblock \emph{\bibinfo{journal}{Phys. Rev.}} \textbf{\bibinfo{volume}{102}},
  \bibinfo{pages}{1217--1230} (\bibinfo{year}{1956}).
\newblock \urlprefix\url{https://link.aps.org/doi/10.1103/PhysRev.102.1217}.

\bibitem{Pershoguba2018}
\bibinfo{author}{Pershoguba, S.~S.} \emph{et~al.}
\newblock Dirac Magnons in Honeycomb Ferromagnets.
\newblock \emph{\bibinfo{journal}{Phys. Rev. X}} \textbf{\bibinfo{volume}{8}},
  \bibinfo{pages}{011010} (\bibinfo{year}{2018}).
\newblock \urlprefix\url{https://link.aps.org/doi/10.1103/PhysRevX.8.011010}.

\end{thebibliography}

\end{document}